%% file: main_submitted.tex
\documentclass[conference]{IEEEtran}
\IEEEoverridecommandlockouts
\usepackage{algorithmic}
\usepackage{amsmath,amssymb,amsfonts}
\usepackage{balance}
\usepackage{booktabs}
\usepackage{cite}
\usepackage{diagbox}
\usepackage{graphicx}
\usepackage{hyperref}
\usepackage{makecell}
\usepackage{multirow}
\usepackage{paralist}
\usepackage{pgfplots}

\pgfplotsset{compat=1.10}
\pgfplotsset{
    discard if/.style 2 args={
        x filter/.code={
            \edef\tempa{\thisrow{#1}}
            \edef\tempb{#2}
            \ifx\tempa\tempb
                \def\pgfmathresult{inf}
            \fi
        }
    },
    discard if not/.style 2 args={
        x filter/.code={
            \edef\tempa{\thisrow{#1}}
            \edef\tempb{#2}
            \ifx\tempa\tempb
            \else
                \def\pgfmathresult{inf}
            \fi
        }
    }
}
\usepackage{subcaption}
\usepackage{textcomp}
\usepackage{url}
\usepackage{verbatim}
\usepackage{xcolor}
\usepackage{xspace}

\def\BibTeX{{\rm B\kern-.05em{\sc i\kern-.025em b}\kern-.08em
    T\kern-.1667em\lower.7ex\hbox{E}\kern-.125emX}}

\makeatletter 
    \newcommand{\linebreakand}{%
    \end{@IEEEauthorhalign}
    \hfill\mbox{}\par
    \mbox{}\hfill\begin{@IEEEauthorhalign}
}
\makeatother 

\newcommand{\mli}[1]{\ensuremath{\mathit{#1}}}
\newcommand{\groundtruth}{\ensuremath{gt}\xspace}
\newcommand{\mediapipe}{\ensuremath{\mathit{Mediapipe}}\xspace}
\newcommand{\holistic}{\ensuremath{\mathit{Holistic}}\xspace}
\newcommand{\FLIC}{\ensuremath{\texttt{FLIC}}\xspace}
\newcommand{\PHOENIX}{\ensuremath{\texttt{PHOENIX}}\xspace}
\newcommand{\FLICShort}{\ensuremath{\texttt{FL}}\xspace}
\newcommand{\PHOENIXShort}{\ensuremath{\texttt{PH}}\xspace}
\newcommand{\MetTest}{\ensuremath{\mli{MT}}\xspace}

\newcommand{\err}[2]{\ensuremath{{\mli{Err^\mli{#2}_{#1}}}}\xspace}
\newcommand{\allRels}{\mli{AllRels}\xspace}
\newcommand{\exampleRels}{\mli{SubRels}\xspace}
\newcommand{\proposed}{\textsc{MeT-Pose}\xspace}
\newcommand{\metRule}{\ensuremath{\mathcal{M}}\xspace}
\newcommand{\transformationName}{\ensuremath{\mathit{trans}}\xspace}
\newcommand{\transformation}{\ensuremath{\mathit{\metRule.\transformationName}}\xspace}
\newcommand{\relationName}{\ensuremath{\mathit{rel}}}
\newcommand{\relation}{\ensuremath{\mathit{\metRule.\relationName}}\xspace}
\newcommand{\SubRate}[2]{\ensuremath{\mathtt{SubRate_{#1, #2}}}}

\newcommand{\Identity}{\ensuremath{\mathtt{Id}}\xspace}
\newcommand{\Stretch}{\ensuremath{\mathtt{Stch}}\xspace}
\newcommand{\StretchArg}[2]{\ensuremath{\Stretch_{#1, #2}}\xspace}
\newcommand{\Mirror}{\ensuremath{\mathtt{Mirr}}\xspace}
\newcommand{\MirrorArg}[2]{\ensuremath{\Mirror_{#1}^{#2}}\xspace}
\newcommand{\Rotation}{\ensuremath{\mathtt{Rot}}\xspace}
\newcommand{\RotationArg}[2]{\ensuremath{\Rotation_{#1}^{#2}}\xspace}
\newcommand{\Bilateral}{\ensuremath{\mathtt{Bilat}}\xspace}
\newcommand{\BilateralArg}[2]{\ensuremath{\Bilateral_{#1}^{#2}}\xspace}
\newcommand{\Motion}{\ensuremath{\mathtt{Motion}}\xspace}
\newcommand{\MotionArg}[2]{\ensuremath{\Motion_{#1}^{#2}}\xspace}
\newcommand{\Resolution}{\ensuremath{\mathtt{Res}}\xspace}
\newcommand{\ResolutionArg}[1]{\ensuremath{\Resolution_{#1}}\xspace}
\newcommand{\Brightness}{\ensuremath{\mathtt{Bright}}\xspace}
\newcommand{\BrightnessArg}[2]{\ensuremath{\Brightness_{#1}^{#2}}\xspace}
\newcommand{\GammaBright}{\ensuremath{\mathtt{Gamma}}\xspace}
\newcommand{\GammaArg}[1]{\ensuremath{\GammaBright_{#1}}\xspace}
\newcommand{\Greyscale}{\ensuremath{\mathtt{Grey}}\xspace}
\newcommand{\Colourwheel}{\ensuremath{\mathtt{CWheel}}\xspace}
\newcommand{\ColourwheelArg}[1]{\ensuremath{\Colourwheel_{#1}}\xspace}
\newcommand{\Colourfill}{\ensuremath{\mathtt{Cfill}}\xspace}
\newcommand{\ColourfillArg}[1]{\ensuremath{\Colourfill_{#1}}\xspace}
\newcommand{\Colourchannels}{\ensuremath{\mathtt{Cchans}}\xspace}
\newcommand{\ColourchannelsArg}[2]{\ensuremath{\Colourchannels_{#1}^{#2}}\xspace}
\newcommand{\Errmetric}[2]{\ensuremath{\mathtt{Err}(#1, #2)}\xspace}
\newcommand{\Masked}[2]{\ensuremath{\mathtt{Flt(#2, #1)}}}
\input{heatMapPreap}

\begin{document}

\title{Metamorphic Testing for Pose Estimation Systems\thanks{This work was supported in part with the financial support of grant 13/RC/2094\_2 to Lero - the Research Ireland Research Centre for Software.}}

\author{\IEEEauthorblockN{Matías Duran \IEEEauthorrefmark{1}}
\and
\IEEEauthorblockN{Thomas Laurent \IEEEauthorrefmark{1}}
\and
\IEEEauthorblockN{Ellen Rushe \IEEEauthorrefmark{2}}
\and
\IEEEauthorblockN{Anthony Ventresque \IEEEauthorrefmark{1}}

\linebreakand
\IEEEauthorblockA{\IEEEauthorrefmark{1} \textit{SFI Lero \& School of Computer Science and Statistics} \\
\textit{Trinity College Dublin} - Dublin, Ireland \\
\{mduran $|$ tlaurent $|$ anthony.ventresque\}@tcd.ie}
\and
\IEEEauthorblockA{\IEEEauthorrefmark{2} \textit{School of Computing} \\
\textit{Dublin City University} - Dublin, Ireland \\
ellen.rushe@dcu.ie}
}

\maketitle

\begin{abstract}

Pose estimation systems are used in a variety of fields, from sports analytics to livestock care. Given their potential impact, it is paramount to systematically test their behaviour and potential for failure. This is a complex task due to the oracle problem and the high cost of manual labelling necessary to build ground truth keypoints. This problem is exacerbated by the fact that different applications require systems to focus on different subjects (e.g., human versus animal) or landmarks (e.g., only extremities versus whole body and face), which makes labelled test data rarely reusable. To combat these problems we propose \proposed, a metamorphic testing framework for pose estimation systems that bypasses the need for manual annotation while assessing the performance of these systems under different circumstances. \proposed thus allows  users of pose estimation systems to assess the systems in conditions that more closely relate to their application without having to label an ad-hoc test dataset or rely only on available datasets, which may not be adapted to their application domain. While we define \proposed in general terms, we also present a non-exhaustive list of metamorphic rules that represent common challenges in computer vision applications, as well as a specific way to evaluate these rules. We then experimentally show the effectiveness of \proposed by applying it to Mediapipe Holistic, a state of the art human pose estimation system, with the \FLIC and \PHOENIX datasets. With these experiments, we outline numerous ways in which the outputs of \proposed can uncover faults in pose estimation systems at a similar or higher rate than classic testing using hand labelled data, and show that users can tailor the rule set they use to the faults and level of accuracy relevant to their application.

\end{abstract}


\input{introduction}
\input{background}

\section{MEtamorphic TEsting for POSe Estimation (\proposed)}
\label{sec:contribution}

This section first formally defines the problem of testing pose estimation systems (\ref{sec:pbDef}) and then introduces \proposed (\ref{sec:approach}).

\subsection{Problem definition}
\label{sec:pbDef}

    \proposed aims to test pose estimation systems at large. Given an input image $img$, a pose estimation system returns an output $O$ which is a list of sets of keypoints: $O = [KP_0, KP_1, ..., KP_n]$ with $n\in\mathbb{N}$ or $O = []$. Each list of keypoints $KP_i$ corresponds to the keypoints of a subject detected in the image by the system, i.e., $\forall i \in \{0..n\}, KP_i = {kp_0, ..., kp_p}$, with each keypoint $kp_j$, identifying a landmark of subject $i$ and its coordinates.

    Testing these systems is an inherently complex task, as they often lack exact requirements. For example, the list of landmarks that should be found on a person in the task of human pose estimation is not defined in a consistent manner with each system defining its own. Similarly, the way to account for occlusion in an image or to process images containing no or multiple subjects is often not defined. This lack of standard requirements contributes to the oracle problem and to the cost of testing pose estimation systems, as each system requires its own hand labelled data for evaluation.

    Additionally, current systems that rely on pose estimation are mostly built using deep neural networks-based pose estimation system. These models are typically trained on large, diverse datasets to facilitate generalisation. Given the ``generalised" nature of these models and high cost of labelling pose data, these models are typically  used ``out-of-the-box", without fine-tuning them for the particular application of the system they are used in, and even sometimes without domain-specific testing. Additionally, different use cases and environmental factors can challenge a pose estimation system in different ways, for example: low lighting, different camera angles, or motion blur. We consider that the problem of testing pose estimation systems should be embedded in a particular use case and consider these application-specific conditions.   

    \proposed aims at tackling these problems with the testing of pose estimation systems. It aims at providing a general, flexible, and tunable testing method for pose estimation systems that does not require hand labelled ground truth test data. \proposed thus offers a testing method that lets users tests these systems for their particular use cases.

\subsection{Approach overview}
\label{sec:approach}
We aim for \proposed to be a general framework so we present it here in broad terms. A specific example of how the framework can be applied is detailed in Section~\ref{sec:metrules}. \proposed is a metamorphic testing framework for pose estimation systems. It thus relies on \emph{metamorphic rules} based on images and keypoints. Each rule \metRule is defined by two elements:
\begin{itemize}
\item A \emph{transformation} \transformation, that defines a modification to apply to an image, e.g., making the image greyscale. Given an original test image $img_{orig}$, \transformation produces a modified test image $img_{mod} = \transformation(img_{orig})$.
\item A \emph{relation} \relation, that defines a property between the System Under Test (SUT)'s output keypoints for $img_{orig}$ and its output keypoints for $img_{mod}$ that should appear if the system functions correctly. The simplest relation is the identity relation, i.e., the system should output the same keypoints for $img_{orig}$ and $img_{mod}$. If this relation is not followed (i.e., the property does not appear), then the rule is said to be violated.
\end{itemize}

A violation of a metamorphic rule on an image indicates that the pose estimation system is providing incorrect output on $img_{orig}$, on $img_{mod}$, or on both. The severity of the system's error can vary, i.e., the output can be more incorrect in some cases than others. This is something that \proposed takes into account by returning the severity of a violation, which can then be considered by, for example, using different error thresholds to decide if a violation constitutes a test failure.

\begin{figure}[htbp]
\includegraphics[width=\columnwidth]{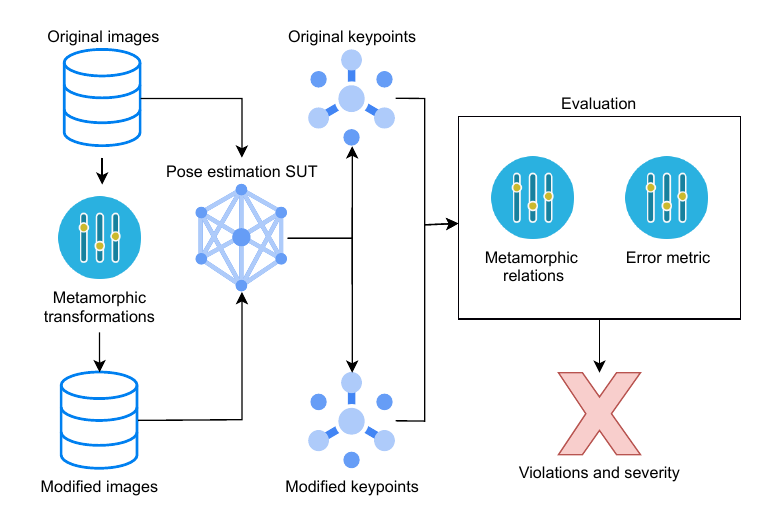}
\caption{Overview diagram of \proposed}
\label{fig:overview}
\end{figure}

Fig.~\ref{fig:overview} gives an overview of the \proposed framework. \proposed takes as input the SUT, a set of metamorphic rules, and a set $IMG_{orig}$ of test images. \proposed applies the transformation of each rule to each test image in $IMG_{orig}$, and for each pair of original and modified images it checks whether the rule is violated, and to what degree. For each input image $img_{orig}\in IMG_{orig}$, and each metamorphic rule \metRule in the input set, the output of \proposed is: 
\begin{inparaitem}
\item a modified image $\transformation(img_{orig})$,
\item whether the pair of images violates \relation, 
\item how severely the pair violates \relation if it does.
\end{inparaitem}
Users are then free to analyse this output based on the quality constraints of their application. Some use cases, for example, might not require that small violations be considered failures if high precision is not required.

\proposed is agnostic to the way rules are defined and how the relations are assessed. The next section details examples of rules and a possible error metric-based technique to assess their relations.

\section{Proposed metamorphic rules}
\label{sec:metrules}

\proposed can be used with any metamorphic rule that defines a transformation \transformationName of an input image $img$ and a relation \relationName \ between the keypoints of $img$ and $\transformationName(img)$. This section describes a non-exhaustive set of rules that we propose to evaluate \proposed with. Section~\ref{sec:proposedRules} describes the transformations and relations of these rules and Section~\ref{sec:metric} defines a possible error-metric based mechanism to assess violations of the rules. Section~\ref{subsec:expError} defines the exact metric used in this work.

\subsection{Transformations and Relations}
\label{sec:proposedRules}

This section describes a varied set of rules for \proposed that cover a range of transformations that are often encountered in the applications of pose estimation systems. In each different application of pose estimation, it can be useful to compare the effect of various modifications to the input images to infer which specific modifications lead to higher output inaccuracies. This is particularly important when there is a possibility of collecting new data for a given task, as these data quality issues can be potentially mitigated or minimised. 

\begin{figure}[t]
    \centering
    \includegraphics[width=\columnwidth]{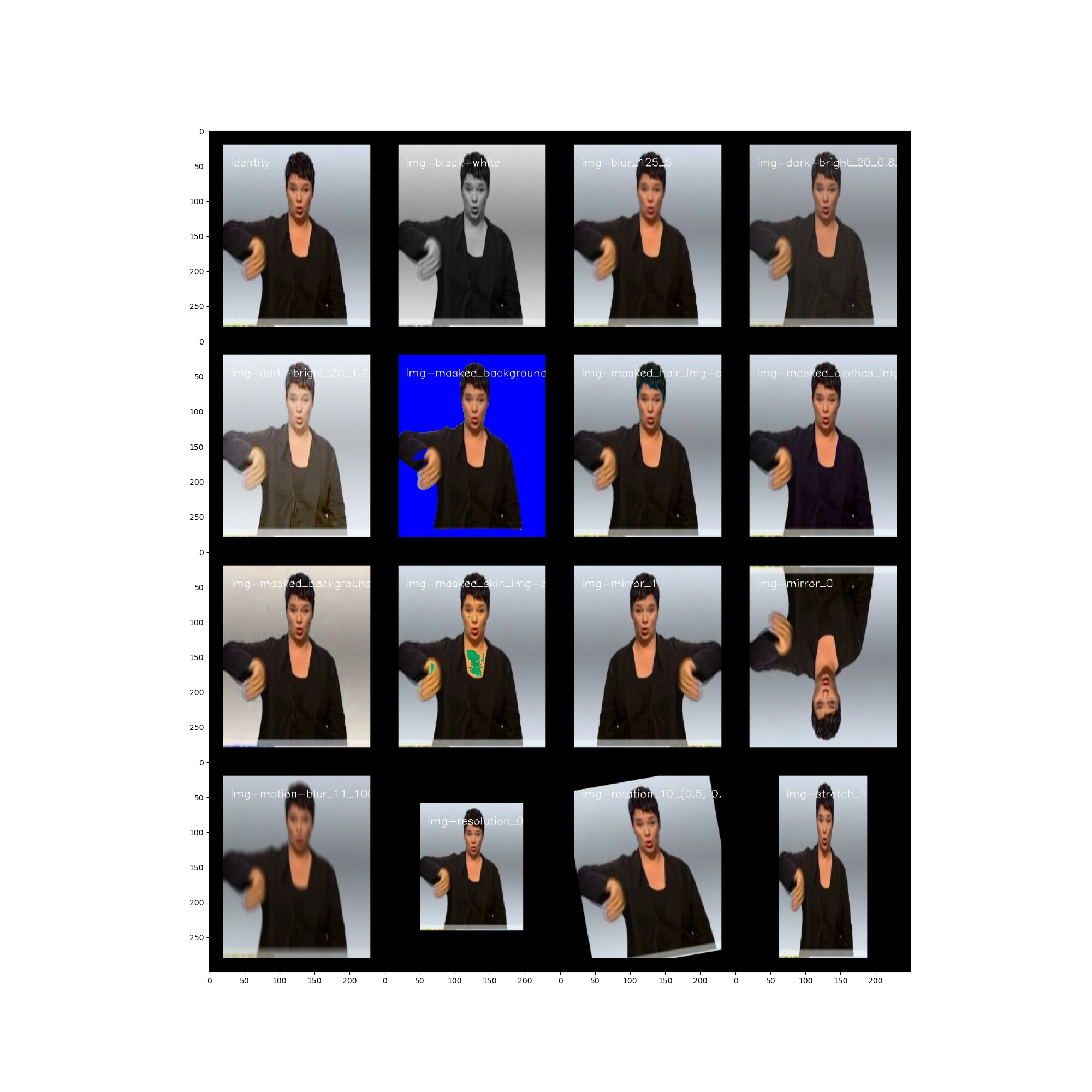}
    \caption{Example image modified by various rules}
    \label{fig:modif_imgs}
\end{figure}

\begin{figure}[!t]
\begin{subfigure}[t]{.33\columnwidth}
\centering
\includegraphics[width=\linewidth]{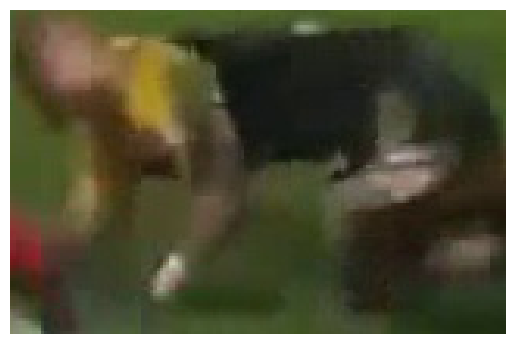}
\label{fig:dog_blur}
\end{subfigure}%
~%
\begin{subfigure}[t]{.33\columnwidth}
\centering
\includegraphics[width=\linewidth]{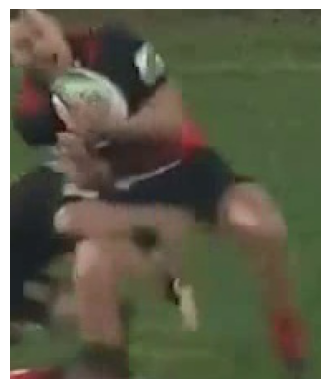}
\label{fig:occlusion_ex}
\end{subfigure}%
~%
\begin{subfigure}[t]{.33\columnwidth}
\centering
\includegraphics[width=\linewidth]{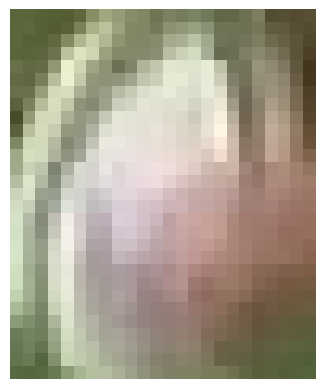} 
\label{fig:pixelated_frisbee}
\end{subfigure}%
\caption{Example inputs from a sports analytics application, showing motion blur and rotation; occlusion; and pixelation}
\label{fig:rugby_examples}
\end{figure}

Fig.~\ref{fig:modif_imgs} shows the effect of some of the transformations used in this work. We detail here the proposed rules and settings used in the experiment. The rules can be classified based on their transformations:

\begin{itemize}
\item Spatial transformations. 

These rules focus on the shape or orientation of the image:
\begin{itemize}
    \item \textbf{Identity} (\Identity): does not modify the image. \relationName: the keypoints on both images should be equal. Checks that the SUT is deterministic.
    \item \textbf{Stretch} (\StretchArg{h}{w}): stretch the image both on the vertical (by a factor of $h$) and horizontal ($w$) axes. \relationName: the keypoints should keep at the same relative coordinates (0,0 is at bottom left of the image, and 1,1 at top right) in the stretched and original images.
    \item  \textbf{Mirror} (\Mirror): mirrors the image along the horizontal (\MirrorArg{h}{}), vertical (\MirrorArg{}{v}) or both (\MirrorArg{h}{v}) axes. \relationName: keypoints should be mirrored in the chosen axes. Mimics situations such as video-calls where the image can be mirrored.
    \item \textbf{Rotation} (\RotationArg{\omega}{c}): rotates the image $\omega$ degrees around a given centre point $c$. \relationName: keypoints on the modified image should be equal to the keypoints on the original image rotated $\omega$ degrees around $c$.
\end{itemize}
\item Image quality transformations.

These rules focus on problems often encountered with the quality of images provided as input to pose estimation systems and challenges often encountered in practice. Fig.~\ref{fig:rugby_examples} provides an example of such problems (motion blur and pixelation) for sports analytics. 

The relation for these rules is that the keypoints should be unchanged after the transformation.
\begin{itemize}
    \item \textbf{Resolution} (\ResolutionArg{factor}): multiplies the resolution of the image by a factor $0<=factor<1$. Lower resolutions of images could be a concern for applications using older or cheaper image capturing devices.
    \item \textbf{Gamma correction} (\GammaArg{\gamma}): applies gamma correction~\cite{gammaCorrection} to the image with a gamma $\gamma$. Brightness can present high variations in real world applications.
    \item \textbf{Brightness Scaling} (\BrightnessArg{a}{m}): for each pixel and each channel value $v^{c}_{i, j}$ returns $a + m * v^{c}_{i, j}$. 
    \item \textbf{Bilateral Filtering} (\BilateralArg{str}{size}): applies a bilateral filter~\cite{bilateralFiltering} of strength ($str$) and size ($size$) to the image, preserving edges of the image but reducing textures. By preserving the edges of objects but reducing textures, this rule checks if the SUT is too reliant on textures instead of the shape of objects, as shown by Geirhos et al.~\cite{geirhos2022imagenettrainedcnnsbiasedtexture}.
    \item \textbf{Motion} (\MotionArg{k\_s}{dir}): applies motion blur to the image with a given kernel size and direction. Motion blur is often seen on images used for pose estimation, as Fig.~\ref{fig:rugby_examples} shows. This is especially true when pose estimation is used for hands~\cite{de2020sign}.
\end{itemize}
\item Colour-space transformations.

These rules change the colours of the image in a way that does not affect the clarity of the subjects in order to ensure that the pose estimation system is properly recognising their pose, and not relying on features unrelated to the pose, such as colours. 

The relation for these rules is that the keypoints should be unchanged after the transformation. All colour transformations test the SUT's over-reliance on colour, which can affect performance when the system is used in a new context~\cite{holmes2022improving}.
\begin{itemize}
    \item \textbf{Greyscale} (\Greyscale): turns the image into a greyscale (only shades of grey) image.
    \item \textbf{Colour wheel} (\ColourwheelArg{\theta}): changes colours in the image by applying a rotation $\theta$ on the hue colour wheel value of a \texttt{HSV}~\cite{HSV} encoding of the input image.
    \item \textbf{Colour channels} (\ColourchannelsArg{[factor1, factor2, factor3]}{encoding}): multiplies each of the 3 colour channels in the $encoding$ (RGB, BGR or XYZ) of the image by a given factor.
\end{itemize}
\end{itemize}

To understand the influence of different zones of the image, we introduce ``filtered'' versions of certain rules: \Masked{zone}{rule}.
The filtered version of a rule only applies the transformation to a particular zone of the image, e.g., the background of the image (all non-subject zones).

The last rule we propose in this work is the \textbf{Colour fill} (\ColourfillArg{colour}). This rule is only used in a filtered way, and therefore fills only a segment of the image a certain colour. Its relation is that keypoints should remain unchanged.

\subsection{Error Metric-based Relation Assessment}
\label{sec:metric}

    This section introduces an error metric-based technique to assess violations of rules used in \proposed. All the rules proposed in this work either expect no change in keypoints or modify the position of the expected output keypoints between the original and modified. However, different rules, e.g., those involving removing or adding landmarks to the image, could also modify the keypoints that the system is expected to detect. Assessing the degree of violation of these different types of rules would require different approaches, which could all be implemented as an error metric  (e.g., counting the number of added landmarks that were not detected by the SUT).

For each rule, the user can provide a metric \Errmetric{O_{expected}}{O_{mod}}, where $O_{expected}$ is the expected position of the keypoints on the modified image according to the metamorphic rule's relation. This metric should be tailored to each relation and to the types of errors considered (e.g., focusing only on hand keypoints). \proposed lets the user define an error threshold $t_{err}$ over which the error is not acceptable. A test then passes iff:
    \[\Errmetric{O_{expected}}{O_{mod}} < t_{err}\]
    
    Both the error metric and error threshold are definable by the user, giving them full control over the aspects of the system's output they want to test (e.g., ignoring particular keypoints), along with the sensitivity of those tests. Section~\ref{subsec:expError} details the human pose estimation-oriented metric used in the experiments, and a comparison of results for different error thresholds.

\section{Experiments}
\label{sec:exps}

This section details the experiments conducted to illustrate how \proposed can be used and evaluate how well it can find faults in a pose estimation system without using ground truth keypoints. Section~\ref{sec:RQs} details the research questions we explore with these experiments. Sections~\ref{subsec:expSUT} and \ref{subsec:expDatasets} detail the pose estimation system under test and datasets used in the experiments, and Section~\ref{subsec:expError} defines the error metric used to evaluate rule violations.

\subsection{Research questions}
\label{sec:RQs}

The experiments in this work aim to answer the following research questions (RQs):

\begin{itemize}
    \item{\bf RQ1}: Does \proposed find faults?

    This RQ considers whether \proposed can find faults in pose estimation systems. In order to explore this question we consider the number of rule violations found by \proposed using different subsets of the rules described in Section~\ref{sec:proposedRules} and different error thresholds.
    
    \item{\bf RQ2}: How different are the results of \proposed and classic, ground truth-based testing?

    This RQ focuses on whether the faults found by \proposed correspond to those found using classic, ground truth-based testing using the same error metric and error threshold as \proposed. This research question considers the same rule sets and error thresholds as RQ1 and is composed of three sub-RQs:
            \begin{itemize}
                \item{\bf RQ2.1}: Do both testing methods lead to a similar number of failures?
                
                    This RQ first assesses whether both testing methods uncover the same number of failing test cases, a test case being considered as an input image for both methods.
                \item{\bf RQ2.2}: Do the same test images make the system fail with both methods?

                    This RQ explores whether the particular images that lead to a failure are the same for both methods, i.e., if they expose the same system failures.
                
                \item{\bf RQ2.3}: How large are the errors found by both methods?

                     Finding larger (w.r.t. the chosen error metric) errors in addition to small errors would show that a testing method finds faults in the SUT with a stronger effect on the output.
            \end{itemize}
            
    \item{\bf RQ3}: What do the different proposed metamorphic rules bring to the method?

        This RQ explores the contribution of the different rules proposed in Section~\ref{sec:proposedRules}\footnote{Note that these rules are general examples that cover general (human) pose estimation challenges and that, as described in Section~\ref{sec:approach}, \proposed can be used with any metamorphic rules, allowing users to tailor the rules they use to the particular application they want to test pose estimation for.} to \proposed. It is composed of two sub-RQs:
        \begin{itemize}
            \item{\bf RQ3.1} Are there subsumption relationships amongst the proposed rules? 

            This RQ explores possible subsumption relations between the different proposed rules. A rule subsumes another rule if it finds the failure-inducing images another rule does, making the second rule redundant.

            \item{\bf RQ3.2} Do the different metamorphic relations reveal the same types of faults?

            This RQ assesses whether the different metamorphic rules find the same failure-inducing images, which would indicate the violations of these rules could have their roots in the same fault in the system.

            \end{itemize}
\end{itemize}

\subsection{System Under Test}\label{subsec:expSUT}

In order to explore the three RQs defined in Section~\ref{sec:RQs}, we applied \proposed to \mediapipe \holistic~\cite{researchMediaPipeHolistic}, a state of the art human pose estimation system. Its underlying DL model is BlazePose GHUM 3D~\cite{grishchenko2022blazepose}. The BlazePose~\cite{bazarevsky2020blazepose} model is a lightweight convolutional neural network that predicts the co-ordinates of body parts.

\holistic is composed of three sub-models, each one responsible for the keypoints of the: face (468 landmarks), hands (21 landmarks each), and overall body pose (33 landmarks).

 We used \holistic in static image mode. Using the system in this mode provides a deterministic output, as verified by the use of the \Identity rule (i.e. there is no difference between two different generations without a transform). This ensures that no noise from non-determinism is integrated into the error metric.

\subsection{Datasets}\label{subsec:expDatasets}

We tested the SUT with \proposed with two datasets: 

\begin{itemize}
    \item \PHOENIX~\cite{phoenixT} is a dataset created for sign language recognition. It contains $947,756$ video frames from a signed German weather forecast. Given that this dataset is used in the field of sign-language recognition and translation, the dataset contains manually labelled annotations in the form of written text translation of signs, but no ground truth keypoints. The images are in a very controlled environment, with all video frames having a similar plain background, with subjects wearing plain black clothes, and without large changes in the subject's distance from the camera or in the angles of their poses. 
    \item \FLIC~\cite{modec13} is a dataset containing $4,552$ movie frames curated from popular movies. All images include one or multiple subjects (people) as it is specifically a human pose estimation benchmark. The environment of this dataset is less controlled, with situations changing drastically depending on the movie scene images were taken from. This provides a contrast to the controlled environment of \PHOENIX. \FLIC contains ground truth (\groundtruth) manual annotations of 11 keypoints obtained from multiple annotators through Amazon Mechanical Turk. We have mapped a subset of keypoints from \mediapipe to the corresponding keypoints of the \FLIC \groundtruth. 
\end{itemize}

Both of these datasets use publicly available images. \FLIC was specifically curated to benchmark pose estimation systems, while Phoenix requires that the person signing remains in frame, meaning that both datasets have an easily identifiable person in the frame. This is in contrast with other datasets that are built for general object classification such as \texttt{COCO}~\cite{COCO}, which contain comparatively fewer human subjects or have these subjects positioned very far away from the camera. We ran our experiments on representative subsets of these two datasets in order to reduce the cost of our experiments. Specifically, for \PHOENIX we used the \textit{dev} subset ($55,775$ images). For \FLIC, we used the images labelled as \texttt{test}, excluding the \textit{FLIC-full} extension, i.e., $835$ images.

\subsection{Metamorphic Rules and Error Metrics}\label{subsec:expError}

In order to explore the research questions that Section~\ref{sec:RQs} details, we applied \proposed using a set of the rules that Section~\ref{sec:proposedRules} describes using a range of configurations of these relations. Table~\ref{tab:metTransforms} details this set of configurations, which is denoted by \allRels.
\begin{table*}[tbp]
    \centering
    \caption{Full list of metamorphic rules settings, settings in bold are included in \exampleRels}
    \label{tab:metTransforms}
    \begin{tabular}{lcp{0.75\textwidth}}
        \toprule
         \textbf{Transformation} & $\in \exampleRels$ & \textbf{Configurations} \\
         \midrule
         \Identity & \checkmark &\\
         \StretchArg{1}{2} & \checkmark & (0.6, 1), (0.8, 1), (0.9, 1.1), (0.95, 1.05), (1, 1.4), (1, 1.25), \textbf{(1, 0.8)}, \textbf{(1, 0.6)}, (1.05, 0.95), (1.1, 0.9), \textbf{(1.25, 1)}, (1.4, 1) \\
         \MirrorArg{h}{v} & \checkmark & \textbf{horizontal}, vertical, both \\  
         \RotationArg{1}{2} & \checkmark & \textbf{(5, (0.5, 0.5))}, \textbf{(10, (0.5, 0.5))}, (15, (0.5, 0.5)), (25, (0.5, 0.5)) \\
         \ResolutionArg{1} & \checkmark & 0.1, \textbf{0.2}, 0.3, 0.4, 0.5, 0.6, \textbf{0.7}, 0.8, 0.9, 0.95, 0.98 \\
         \GammaArg{1} & \checkmark & 0.25, \textbf{0.5}, 0.85, 0.95, 1.05, 1.15, 1.5, 1.75 \\
         \BrightnessArg{1}{2} & \checkmark & (-20, 0.8), (-20, 1.6), (0, 1.05), (0, 1.15), (20, 0.4), \textbf{(20, 0.8)}, (20, 1.2), (20, 1.6), (30, 1.15)  \\
         \BilateralArg{1}{2} & \checkmark & \textbf{(10, 3)}, (10, 5), (10, 7), (10, 9), (30, 3), (30, 5), (30, 7), (30, 9), (50, 3), (50, 5), (50, 7), (50, 9), (80, 3), (80, 5), \textbf{(80, 7)}, (80, 9), (125, 3), \textbf{(125, 5)}, (125, 7), (125, 9), (150, 3), (150, 5), (150, 7), (150, 9), (180, 3), (180, 5), (180, 7), (180, 9)  \\
         \MotionArg{1}{2} & \checkmark & (5, 0), (5, 40), (5, 70), (5, 100), (7, 0), (7, 40), (7, 70), (7, 100), (9, 0), (9, 40), (9, 70), (9, 100), \textbf{(11, 0)},  (11, 70), \textbf{(11, 100)} \\
         \Greyscale & \checkmark \\
         \Masked{2}{\ColourwheelArg{1}} & \checkmark & (10, skin), (30, skin), (90, skin), (-45, skin), (10, clothes), (30, clothes), (90, clothes), (-45, clothes), \textbf{(90, hair)}, \textbf{(90, background)} \\
         \Masked{3}{\ColourchannelsArg{1}{2}} & & ([0.9, 1.1, 1.1], RGB, skin), ([1.1, 1.1, 0.9], RGB, skin), ([0.8, 1.3, 1.3], RGB, skin), ([1.3, 1.3, 0.8], RGB, skin), ([0.6, 1.4, 1], RGB, skin), ([1.4, 1, 0.6], RGB, skin), ([0.45, 1, 1.2], RGB, skin), ([1.2, 1, 0.45], RGB, skin), ([1, 1, 1], BGR, skin), ([1, 1, 1], XYZ, skin) \\
         \Masked{2}{\ColourfillArg{1}} &  & ([0, 0, 255], background), ([255, 180, 120]$^a$, background), ([33, 28, 27]$^b$, background), ([0, 0, 255], skin), ([255, 180, 120], skin), ([33, 28, 27], skin), ([0, 0, 255], clothes), ([255, 180, 120], clothes), ([33, 28, 27], clothes) \\
         \bottomrule
    \end{tabular}
    \flushright{\footnotesize\flushright
$^a$ close to skin colour on phoenix dataset \hspace{2mm} $^b$close to clothes colour in phoenix dataset}
\end{table*}

Not all of the relations and configurations in \allRels are relevant to the data in \PHOENIX and \FLIC and to testing \holistic without finetuning, leading to artificially high error rates. We thus also use a restricted set of relations and settings, highlighted in bold in Table~\ref{tab:metTransforms} and denoted by \exampleRels. Additionally, we consider \Greyscale and \MirrorArg{h}{} in order to assess whether \proposed can find problems with \textit{concept understanding} in these systems -- an area previously shown to be faulty in other deep learning-based models~\cite{hoehing2023s}. Finally, we determine whether \proposed can surface particular faults in the pose estimation system such as over-reliance on colour (which can vary substantially based on the lighting and other conditions) or orientation (e.g., dealing with left and right handed signers in \PHOENIX, or with mirrored input from online calls)-- two features that should not impact its output.  

To assess violations of all rules used in the experiments we use the same error metric \err{lmks}{} defined as:
{\footnotesize\[
    \err{lmks}{}(O_{expct}, O_{trans})= 
\begin{cases}
    \text{if } O_{expct} = \emptyset \veebar  O_{trans} = \emptyset, \inf\\
    \text{elif } O_{expct} = \emptyset \land  O_{trans} = \emptyset, 0\\
    \text{else}, \underset{lmk\in lmks}{\text{Med}}L2_{MP}(kp_{expct}, kp_{trans})
\end{cases}
\]}

If the pose estimation system only returns keypoints on the original image or the modified image, \err{lmks}{} returns an infinite value. Indeed, the transformations used in these experiments do not modify which landmarks the system should detect on the image, at most they change the expected position of the returned keypoints. Failing to detect any landmarks on just one of the pair of images is thus the worst violation of a rule possible. 
However, if the pose estimation system returns no keypoints for both images it displays a coherent behaviour w.r.t. the metamorphic rule and the rule is not violated. Note that, in general, not returning keypoints does not denote a fault in itself as an image could contain no visible person. However, \FLIC and \PHOENIX both only contain images containing people so $O_{orig} = \emptyset$ does denote a fault, albeit not one that would violate our proposed rules.

If the pose estimation system returned keypoints for both images then \err{lmks}{} returns the median of the distance between the expected keypoint (following the metamorphic rule's relation and the keypoints returned by the system on the original image) and the keypoint returned by the system on the modified image for each landmark $lmk$.
Using the median as a measure of the overall error of the system on an image attenuates the effect of outliers on the error value. Indeed, if only a single keypoint's coordinates are extremely inaccurate, many applications would still work. Note however that the error of all keypoints in an image could be aggregated in other ways (min, max, \ldots) depending on the needs of the use case the pose estimation system is tested for.
Here the distance used, $L2_{MP}$ is the normalised Euclidean distance as described in \mediapipe's model cards~\cite{modelcard2021mediapipe, modelcardface, modelcardhands}, i.e., depending on which of the 3 groups described in section~\ref{subsec:expSUT} the landmark belongs to, the distance is normalised by dividing by either the distance between the subject's: shoulders; irises; or their wrist and the first joint of their middle finger.

As we test \mediapipe without a particular use case, setting a meaningful value for the error threshold is not possible. We thus assess violations using a large range of values for the threshold and report the performance of \proposed using these different values.

\section{Results}
\label{sec:results}

This section analyses and discusses some of the results from our experiments following the research questions defined in Section~\ref{sec:RQs}. The code used to generate the reported results is available in the companion repository~\cite{publicationRepo}.

\subsection{RQ1}
\label{subsec:rq1}

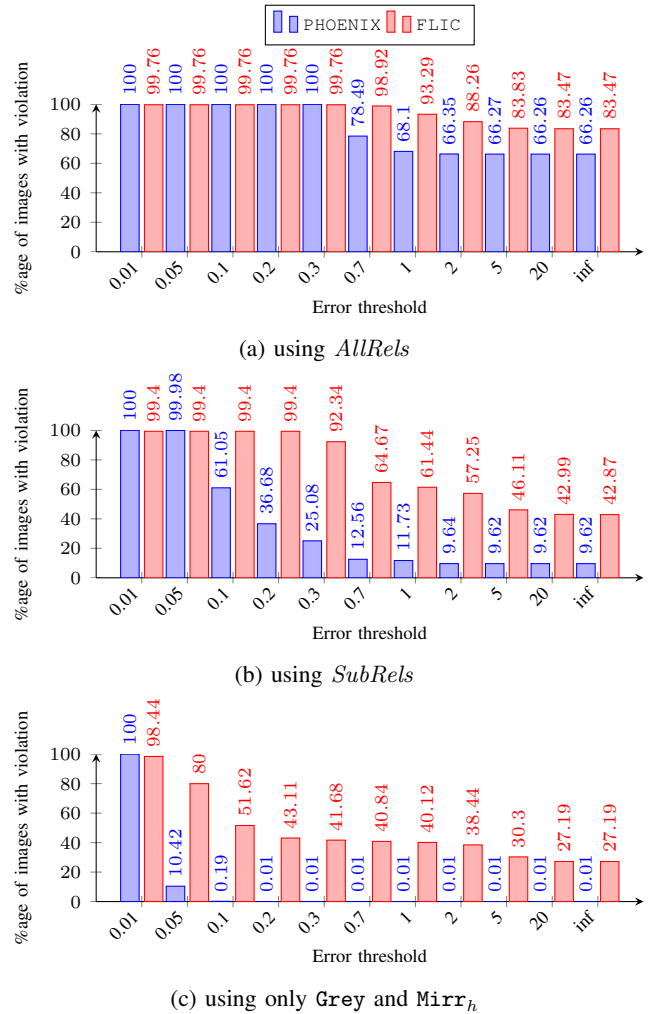
\begin{figure}[tbp]
\begin{subfigure}{\columnwidth}
\centering
\begin{tikzpicture}
\begin{axis}[
width=\columnwidth,
height=0.15\textheight,
ybar,
bar width=7pt,
nodes near coords,
axis lines=left,
enlarge x limits=true,
ymin = 0,
ymax=100,
legend style={at={(0.5, 1.4)}, anchor=south, legend columns=-1, font=\scriptsize},
y tick label style ={font=\small},
ylabel={\%age of images with violation},
ylabel style ={font=\scriptsize},
x tick label style ={font=\small},
xlabel={Error threshold},
xlabel style ={font=\scriptsize, yshift=5},
symbolic x coords={0.01, 0.05, 0.1, 0.2, 0.3, 0.7, 1, 2, 5, 20, inf},
xtick=data,
x tick label style={rotate=45,anchor=east, font=\scriptsize},
y tick label style={font=\scriptsize},
every node near coord/.append style={rotate=90,anchor=west, font=\scriptsize, /pgf/number format/.cd,fixed
,precision=2}
]
\addplot coordinates {(0.01, 100) (0.05, 100) (0.1, 100) (0.2, 100) (0.3, 100) (0.7, 78.49) (1, 68.1) (2, 66.35) (5, 66.27) (20, 66.26) (inf, 66.26)};
\addplot table [x=plot_thresholds, y expr=\thisrow{error_counts} * 100, col sep=comma] {flic_no_gt_all.txt};
\legend{\PHOENIX, \FLIC}
\end{axis}
\end{tikzpicture}
\caption{using \allRels}
\label{fig:results_pose}
\end{subfigure}%

\begin{subfigure}{\columnwidth}
\centering
\begin{tikzpicture}
\begin{axis}[
width=\columnwidth,
height=0.15\textheight,
ybar,
bar width=7pt,
nodes near coords,
axis lines=left,
enlarge x limits=true,
ymin = 0,
ymax=100,
y tick label style ={font=\small},
ylabel={\%age of images with violation},
ylabel style ={font=\scriptsize},
x tick label style ={font=\small},
xlabel={Error threshold},
xlabel style ={font=\scriptsize, yshift=5},
symbolic x coords={0.01, 0.05, 0.1, 0.2, 0.3, 0.7, 1, 2, 5, 20, inf},
xtick=data,
x tick label style={rotate=45,anchor=east, font=\scriptsize},
y tick label style={font=\scriptsize},
every node near coord/.append style={rotate=90,anchor=west, font=\scriptsize, /pgf/number format/.cd,fixed 
,precision=2}
]
\addplot coordinates {(0.01, 100) (0.05, 99.98) (0.1, 61.05) (0.2, 36.68) (0.3, 25.08) (0.7, 12.56) (1, 11.73) (2, 9.64) (5, 9.62) (20, 9.62) (inf, 9.62)};
\addplot table [x=plot_thresholds, y expr=\thisrow{error_counts} * 100, col sep=comma] {flic_no_gt_subRels.txt};
\end{axis}
\end{tikzpicture}
\caption{using \exampleRels}
\label{fig:pose_subrel}
\end{subfigure}%

\begin{subfigure}{\columnwidth}
\centering
\begin{tikzpicture}
\begin{axis}[
width=\columnwidth,
height=0.15\textheight,
ybar,
bar width=7pt,
nodes near coords,
axis lines=left,
enlarge x limits=true,
ymin = 0,
ymax=100,
y tick label style ={font=\small},
ylabel={\%age of images with violation},
ylabel style ={font=\scriptsize},
x tick label style ={font=\small},
xlabel={Error threshold},
xlabel style ={font=\scriptsize, yshift=5},
symbolic x coords={0.01, 0.05, 0.1, 0.2, 0.3, 0.7, 1, 2, 5, 20, inf},
xtick=data,
x tick label style={rotate=45,anchor=east, font=\scriptsize},
y tick label style={font=\scriptsize},
every node near coord/.append style={rotate=90,anchor=west, font=\scriptsize, /pgf/number format/.cd,fixed 
,precision=2}
]
\addplot coordinates {(0.01, 100) (0.05, 10.42) (0.1, 0.19) (0.2, 0.01) (0.3, 0.01) (0.7, 0.01) (1, 0.01) (2, 0.01) (5, 0.01) (20, 0.01) (inf, 0.01)};
\addplot table [x=plot_thresholds, y expr=\thisrow{error_counts} * 100, col sep=comma] {flic_no_gt_bw_mirror.txt};
\end{axis}
\end{tikzpicture}
\caption{using only \Greyscale and \MirrorArg{h}{}}
\label{fig:pose_b&w}
\end{subfigure}%
\caption{Percentage of images leading to a rule violation with varying error thresholds using body landmarks}
\label{fig:pose_results}
\end{figure}

Figure~\ref{fig:pose_results} shows the proportion of images in each dataset leading to a rule violation when testing \holistic with \proposed for different sets of rules and different error threshold values when considering the body pose landmarks. These results confirm that \proposed can find faults in the system as it finds violations of the rules. They also confirm the central role of the rules and the error threshold value used in \proposed, which is why we have chosen to keep these characteristics customisable so they can be defined by users of the framework
to suit their particular application. 

When the threshold is set very low, e.g., to 0.01\footnote{In this setting, a median error of more than 1\% of the distance between the subject's shoulders leads to a violation.} as the leftmost columns of the graphs illustrate, \proposed is very sensitive and most images lead to violations. With high error threshold values \proposed is much more lenient and the proportion of images leading to a violation quickly drops. However, \proposed still finds images that lead to an infinite error, i.e., where the pose estimation system detects the subject only in one of the original and modified images, as defined in Section~\ref{subsec:expError}. These cases are always considered violations of our proposed metamorphic rules.

\begin{figure}[tbp]    
    \begin{subfigure}{\columnwidth}
    \centering
    \begin{tikzpicture}
\begin{axis}[
    width=\columnwidth,
    height=0.15\textheight,
    ybar,
	bar width=7pt,
	nodes near coords,
    axis lines=left,
    enlarge x limits=true,
    ymin = 0,
    ymax=100,
    legend style={at={(0.5, 1)}, anchor=north,legend columns=-1},
    y tick label style ={font=\small},
    ylabel={\%age of images with violation},
    ylabel style ={font=\scriptsize},
    x tick label style ={font=\small},
    xlabel={Error threshold},
    xlabel style ={font=\scriptsize},
    symbolic x coords={0.01, 0.05, 0.1, 0.2, 0.3, 0.7, 1, 2, 5, 20, inf},
    xtick=data,
    x tick label style={rotate=45,anchor=east},
    every node near coord/.append style={rotate=90,anchor=west, font=\scriptsize, /pgf/number format/.cd,fixed zerofill,precision=2}
    ]
\addplot table [x=plot_thresholds, y expr=\thisrow{error_counts}*100,
col sep=comma, discard if not={feature}{left_hand}] {phoenix_mirror_inf.txt};
\addplot table [x=plot_thresholds, y expr=\thisrow{error_counts} * 100, col sep=comma, discard if={feature}{left_hand}] {phoenix_mirror_inf.txt};
\legend{left hand, right hand}
\end{axis}
\end{tikzpicture}
    \caption{Using only \MirrorArg{h}{}}
    \label{fig:mirror_hands_phoenix}
    \end{subfigure}%
    
    \begin{subfigure}{\columnwidth}
    \centering
    \begin{tikzpicture}
\begin{axis}[
    width=\columnwidth,
    height=0.15\textheight,
    ybar,
	bar width=7pt,
	nodes near coords,
    axis lines=left,
    enlarge x limits=true,
    ymin = 0,
    ymax=100,
    y tick label style ={font=\small},
    ylabel={\%age of images with violation},
    ylabel style ={font=\scriptsize},
    x tick label style ={font=\small},
    xlabel={Error threshold},
    xlabel style ={font=\scriptsize},
    symbolic x coords={0.01, 0.05, 0.1, 0.2, 0.3, 0.7, 1, 2, 5, 20, inf},
    xtick=data,
    x tick label style={rotate=45,anchor=east},
    every node near coord/.append style={rotate=90,anchor=west, font=\scriptsize, /pgf/number format/.cd,fixed zerofill,precision=2}
    ]
    
\addplot table [x=plot_thresholds, y expr=\thisrow{error_counts}*100,
col sep=comma, discard if not={feature}{left_hand}] {phoenix_bw_inf.txt};
\addplot table [x=plot_thresholds, y expr=\thisrow{error_counts} * 100, col sep=comma, discard if={feature}{left_hand}] {phoenix_bw_inf.txt};
\end{axis}
\end{tikzpicture}
    \caption{Using only \Greyscale}
    \label{fig:BW_hands_phoenix}
    \end{subfigure}%
    ~%
    \caption{Percentage of images in \PHOENIX leading to a rule violation for varying error thresholds using hand landmarks}
    \label{fig:phoenix_hands_results}
    \end{figure}
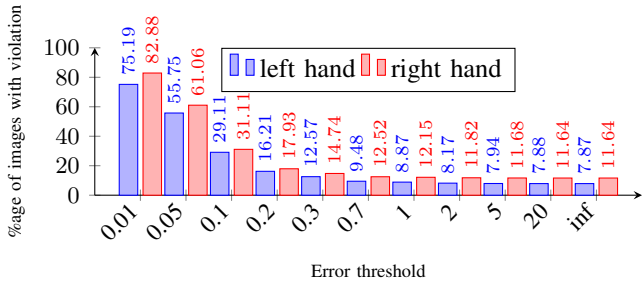
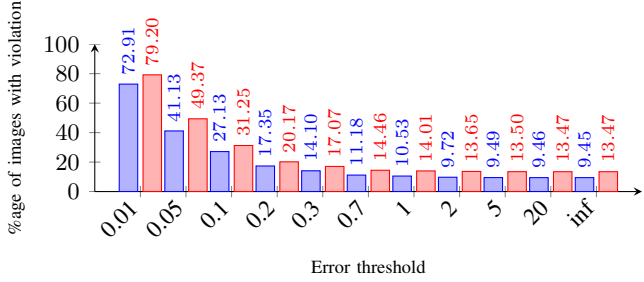

Figure~\ref{fig:phoenix_hands_results} shows the proportion of images in \PHOENIX leading to a rule violation with \Greyscale and \MirrorArg{h}{} for different error threshold values when considering the hand landmarks. These results, when contrasted with those in Figure~\ref{fig:pose_results}, show once more the importance of adaptability in \proposed. While focusing on the body pose landmarks quickly exposes very few errors on \FLIC when using only the \Greyscale and \MirrorArg{h}{}, using hand landmarks exposes many more rule violations. This is consistent with the goal of \PHOENIX, i.e., the fact that it is a sign language dataset, where the hands, arms, and face, are much more the focus than the general body pose.

\subsection{RQ2}

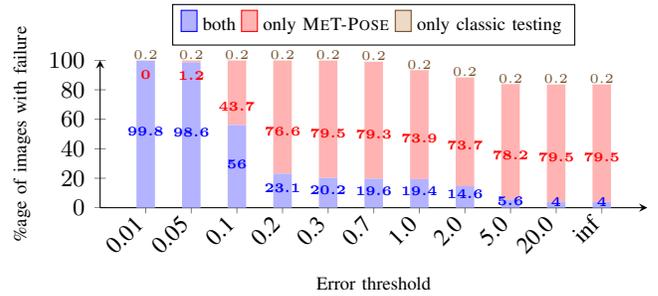
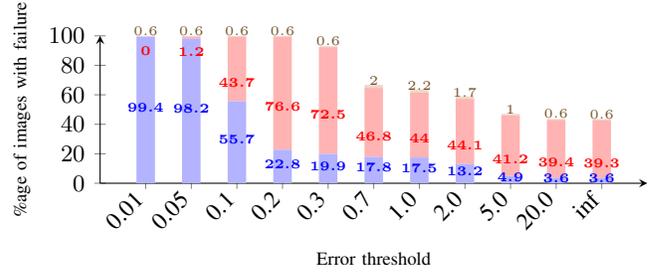
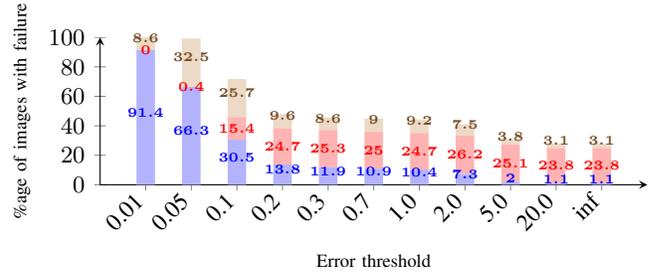
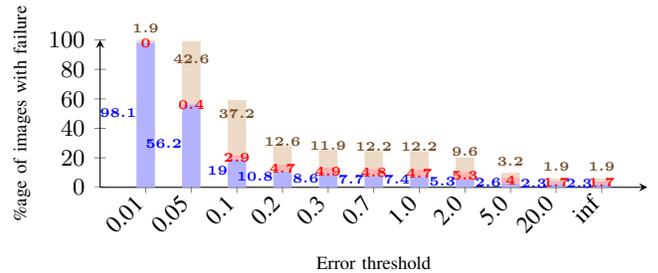
\begin{figure}[tb]
\centering
\begin{subfigure}{\columnwidth}
\begin{tikzpicture}
\begin{axis} [
width=\columnwidth,
height=0.15\textheight,
ybar stacked,
stacked ignores zero=false,
bar width=7pt,
axis lines=left,
enlarge x limits=true,
ymin = 0,
ymax=100,
legend style={at={(0.5, 1.1)}, anchor=south,legend columns=-1, font=\scriptsize},
y tick label style ={font=\small},
ylabel={\%age of images with failure},
ylabel style ={font=\scriptsize},
x tick label style ={font=\small},
xlabel={Error threshold},
xlabel style ={font=\scriptsize},
symbolic x coords={0.01, 0.05, 0.1, 0.2, 0.3, 0.7, 1.0, 2.0, 5.0, 20.0, inf},
xtick=data,
x tick label style={rotate=45,anchor=east},
]
\addplot+ [draw=none, nodes near coords, every node near coord/.append style={font=\tiny\boldmath, yshift=1pt,/pgf/number format/.cd,fixed,precision=1}] table [x=plot_thresholds, y expr=\thisrow{error_counts} * 100, col sep=comma, discard if not={feature}{both failed}] {flic_all_inf.txt};
\addplot+ [draw=none, nodes near coords, every node near coord/.append style={font=\tiny\boldmath, anchor = north,/pgf/number format/.cd,fixed,precision=1}] table [x=plot_thresholds, y expr=\thisrow{error_counts}*100, col sep=comma, discard if not={feature}{only metamorphic}] {flic_all_inf.txt};
\addplot+ [draw=none, nodes near coords, every node near coord/.append style={font=\tiny, anchor = south, yshift=-3,/pgf/number format/.cd,fixed,precision=1}] table [ x=plot_thresholds, y expr=\thisrow{error_counts}*100, col sep=comma, discard if not={feature}{only ground truth}] {flic_all_inf.txt};
\legend{both, only \proposed, only classic testing}
\end{axis}
\end{tikzpicture}
\caption{with \proposed using \allRels}
\end{subfigure}
\begin{subfigure}{\columnwidth}
\begin{tikzpicture}
\begin{axis}[
width=\columnwidth,
height=0.15\textheight,
ybar stacked,
stacked ignores zero=false,
bar width=7pt,
axis lines=left,
enlarge x limits=true,
ymin = 0,
ymax=100,
y tick label style ={font=\small},
ylabel={\%age of images with failure},
ylabel style ={font=\scriptsize},
x tick label style ={font=\small},
xlabel={Error threshold},
xlabel style ={font=\scriptsize},
symbolic x coords={0.01, 0.05, 0.1, 0.2, 0.3, 0.7, 1.0, 2.0, 5.0, 20.0, inf},
xtick=data,
x tick label style={rotate=45,anchor=east},
]
\addplot+ [draw=none, nodes near coords, every node near coord/.append style={font=\tiny\boldmath, yshift=1pt,/pgf/number format/.cd,fixed,precision=1}] table [x=plot_thresholds, y expr=\thisrow{error_counts} * 100, col sep=comma, discard if not={feature}{both failed}] {flic_exRels_inf.txt};
\addplot+ [draw=none, nodes near coords, every node near coord/.append style={font=\tiny\boldmath, anchor = north,/pgf/number format/.cd,fixed,precision=1}] table [x=plot_thresholds, y expr=\thisrow{error_counts}*100, col sep=comma, discard if not={feature}{only metamorphic}] {flic_exRels_inf.txt};
\addplot+ [draw=none, nodes near coords, every node near coord/.append style={font=\tiny, anchor = south, yshift=-3,/pgf/number format/.cd,fixed,precision=1}] table [ x=plot_thresholds, y expr=\thisrow{error_counts}*100, col sep=comma, discard if not={feature}{only ground truth}] {flic_exRels_inf.txt};
\end{axis}
\end{tikzpicture}
\caption{with \proposed using \exampleRels}
\end{subfigure}
\begin{subfigure}{\columnwidth}
\begin{tikzpicture}
\begin{axis}[
width=\columnwidth,
height=0.15\textheight,
ybar stacked,
stacked ignores zero=false,
bar width=7pt,
axis lines=left,
enlarge x limits=true,
ymin = 0,
ymax=100,
y tick label style ={font=\small},
ylabel={\%age of images with failure},
ylabel style ={font=\scriptsize},
x tick label style ={font=\small},
xlabel={Error threshold},
xlabel style ={font=\scriptsize},
symbolic x coords={0.01, 0.05, 0.1, 0.2, 0.3, 0.7, 1.0, 2.0, 5.0, 20.0, inf},
xtick=data,
x tick label style={rotate=45,anchor=east},
]
\addplot+ [draw=none, nodes near coords, every node near coord/.append style={font=\tiny\boldmath, yshift=2pt,/pgf/number format/.cd,fixed,precision=1}] table [x=plot_thresholds, y expr=\thisrow{error_counts} * 100, col sep=comma, discard if not={feature}{both failed}] {flic_bw_inf.txt};
\addplot+ [draw=none, nodes near coords, every node near coord/.append style={font=\tiny\boldmath,/pgf/number format/.cd,fixed,precision=1}] table [x=plot_thresholds, y expr=\thisrow{error_counts}*100, col sep=comma, discard if not={feature}{only metamorphic}] {flic_bw_inf.txt};
\addplot+ [draw=none, nodes near coords, every node near coord/.append style={font=\tiny\boldmath, anchor=south, yshift=-3,/pgf/number format/.cd,fixed,precision=1}] table [ x=plot_thresholds, y expr=\thisrow{error_counts}*100, col sep=comma, discard if not={feature}{only ground truth}] {flic_bw_inf.txt};
\end{axis}
\end{tikzpicture}
\caption{with \proposed using \Greyscale}
\end{subfigure}
\begin{subfigure}{\columnwidth}
\begin{tikzpicture}
\begin{axis}[
width=\columnwidth,
height=0.15\textheight,
ybar stacked,
stacked ignores zero=false,
bar width=7pt,
axis lines=left,
enlarge x limits=true,
ymin = 0,
ymax=100,
y tick label style ={font=\small},
ylabel={\%age of images with failure},
ylabel style ={font=\scriptsize},
x tick label style ={font=\small},
xlabel={Error threshold},
xlabel style ={font=\scriptsize},
symbolic x coords={0.01, 0.05, 0.1, 0.2, 0.3, 0.7, 1.0, 2.0, 5.0, 20.0, inf},
xtick=data,
x tick label style={rotate=45,anchor=east},
]
\addplot+ [draw=none, nodes near coords, every node near coord/.append style={font=\tiny\boldmath, anchor=east, 
yshift=1pt, 
/pgf/number format/.cd,fixed,precision=1}] table [x=plot_thresholds, y expr=\thisrow{error_counts} * 100, col sep=comma, discard if not={feature}{both failed}] {flic_mirror_inf.txt};
\addplot+ [draw=none, nodes near coords, every node near coord/.append style={font=\tiny\boldmath, /pgf/number format/.cd,fixed,precision=1}] table [x=plot_thresholds, y expr=\thisrow{error_counts}*100, col sep=comma, discard if not={feature}{only metamorphic}] {flic_mirror_inf.txt};
\addplot+ [draw=none, nodes near coords, every node near coord/.append style={font=\tiny\boldmath, anchor=south, /pgf/number format/.cd,fixed,precision=1}] table [ x=plot_thresholds, y expr=\thisrow{error_counts}*100, col sep=comma, discard if not={feature}{only ground truth}] {flic_mirror_inf.txt};
\end{axis}
\end{tikzpicture}
\caption{with \proposed using \MirrorArg{h}{}}
\end{subfigure}
\caption{Percentage of images in \FLIC leading to a classic test failure or a rule violation with varying error thresholds using body landmarks}
\label{fig:rq2}
\end{figure}

This RQ focuses on the difference between classic, ground truth based testing and \proposed. Thus, it relies on \FLIC, the only dataset in our experiments with ground truth labels. Figure~\ref{fig:rq2} shows: the proportion of images leading to both a classic test failure and a rule violation, a classic test failure only, and a rule violation only for different error thresholds. A classic test failure is defined as a normalised distance between the ground truth keypoints and the keypoints output by the system on the original image greater than the error threshold.
These results show that for low values of the error threshold, both methods perform similarly in the number of failures they find, even when using a single rule in \proposed. This is explained by both methods being too sensitive with that configuration and considering nearly all outputs as faults. With higher values of the error threshold we see that \proposed finds many more failures than classic testing when using \allRels or \exampleRels. The answer to RQ2.1 is thus that, overall, \proposed finds more failures than classic testing. 

Regarding RQ2.2, we see that in all cases, there is some overlap between the images that lead to failures for each method, however each method also finds failures that the other method did not detect, with \proposed finding more failures when using more complete sets of rules.

As discussed in RQ2.1, both testing methods perform similarly for smaller values of the error threshold while \proposed finds more failures for higher values. This indicates that \proposed finds larger failures overall.

\subsection{RQ3}

\fpSet \heatMapMax {1}
\begin{figure}[!tb]
\centering
\begin{subfigure}{\columnwidth}
\resizebox{\columnwidth}{!}{%
\begin{tblr}{colspec={lccccccccccccccccccc}, process=\funcBgColorDouble, row{2-20}={1.5em}, column{2-20} ={1.5em}, row{1}={cmd={\rotatebox{30}}}}
& \Identity & \Greyscale & \BilateralArg{10}{3} & \BilateralArg{125}{5} & \BilateralArg{80}{7} & \BrightnessArg{20}{0.8} & \GammaArg{0.5} & \Masked{bg}{\ColourwheelArg{90}} & \Masked{hair}{\ColourwheelArg{90}} & \MirrorArg{h}{} & \MotionArg{11}{0} & \MotionArg{11}{100} & \ResolutionArg{0.2} & \ResolutionArg{0.7} & \RotationArg{0.5}{(0.5, 0.5)} & \RotationArg{10}{(0.5, 0.5)} & \StretchArg{1.25}{1} & \StretchArg{1}{0.6} & \StretchArg{1}{0.8}\\
\Identity & 1.0 & 1.0 & 1.0 & 1.0 & 1.0 & 1.0 & 1.0 & 1.0 & 1.0 & 1.0 & 1.0 & 1.0 & 1.0 & 1.0 & 1.0 & 1.0 & 1.0 & 1.0 & 1.0\\
\Greyscale & 0.0 & 1.0 & 0.0 & 0.0 & 0.0 & 0.0 & 0.0 & 0.0 & 0.0 & 0.0 & 0.0 & 0.0 & 0.2 & 0.0 & 0.0 & 0.0 & 0.2 & 0.4 & 0.2\\
\BilateralArg{10}{3} & 1.0 & 1.0 & 1.0 & 1.0 & 1.0 & 1.0 & 1.0 & 1.0 & 1.0 & 1.0 & 1.0 & 1.0 & 1.0 & 1.0 & 1.0 & 1.0 & 1.0 & 1.0 & 1.0\\
\BilateralArg{125}{5} & 0.0 & 0.0 & 0.0 & 1.0 & 0.0 & 0.0 & 1.0 & 0.0 & 0.0 & 0.0 & 0.0 & 1.0 & 0.0 & 0.0 & 0.0 & 0.0 & 0.0 & 1.0 & 0.0\\
\BilateralArg{80}{7} & 1.0 & 1.0 & 1.0 & 1.0 & 1.0 & 1.0 & 1.0 & 1.0 & 1.0 & 1.0 & 1.0 & 1.0 & 1.0 & 1.0 & 1.0 & 1.0 & 1.0 & 1.0 & 1.0\\
\BrightnessArg{20}{0.8} & 1.0 & 1.0 & 1.0 & 1.0 & 1.0 & 1.0 & 1.0 & 1.0 & 1.0 & 1.0 & 1.0 & 1.0 & 1.0 & 1.0 & 1.0 & 1.0 & 1.0 & 1.0 & 1.0\\
\GammaArg{0.5} & 0.0 & 0.0 & 0.0 & 1.0 & 0.0 & 0.0 & 1.0 & 0.0 & 0.0 & 0.0 & 0.0 & 1.0 & 0.0 & 0.0 & 0.0 & 0.0 & 0.0 & 1.0 & 0.0\\
\Masked{bg}{\ColourwheelArg{90}} & 1.0 & 1.0 & 1.0 & 1.0 & 1.0 & 1.0 & 1.0 & 1.0 & 1.0 & 1.0 & 1.0 & 1.0 & 1.0 & 1.0 & 1.0 & 1.0 & 1.0 & 1.0 & 1.0\\
\Masked{hair}{\ColourwheelArg{90}} & 1.0 & 1.0 & 1.0 & 1.0 & 1.0 & 1.0 & 1.0 & 1.0 & 1.0 & 1.0 & 1.0 & 1.0 & 1.0 & 1.0 & 1.0 & 1.0 & 1.0 & 1.0 & 1.0\\
\MirrorArg{h}{} & 0.0 & 0.0 & 0.0 & 0.0 & 0.0 & 0.0 & 0.0 & 0.0 & 0.0 & 1.0 & 0.0 & 0.0 & 0.0 & 0.0 & 0.0 & 0.0 & 1.0 & 1.0 & 1.0\\
\MotionArg{11}{0} & 0.0 & 0.0 & 0.0 & 0.0 & 0.0 & 0.0 & 0.0 & 0.0 & 0.0 & 0.0 & 1.0 & 0.025974025974025976 & 0.37662337662337664 & 0.0 & 0.0 & 0.0 & 0.0 & 0.6233766233766234 & 0.0\\
\MotionArg{11}{100} & 0.0 & 0.0 & 0.0 & 0.030303030303030304 & 0.0 & 0.0 & 0.030303030303030304 & 0.0 & 0.0 & 0.0 & 0.06060606060606061 & 1.0 & 0.2727272727272727 & 0.0 & 0.0 & 0.0 & 0.06060606060606061 & 0.7272727272727273 & 0.06060606060606061\\
\ResolutionArg{0.2} & 0.0 & 0.0034129692832764505 & 0.0 & 0.0 & 0.0 & 0.0 & 0.0 & 0.0 & 0.0 & 0.0 & 0.09897610921501707 & 0.030716723549488054 & 1.0 & 0.0 & 0.0 & 0.0 & 0.006825938566552901 & 0.6279863481228669 & 0.0034129692832764505\\
\ResolutionArg{0.7} & 1.0 & 1.0 & 1.0 & 1.0 & 1.0 & 1.0 & 1.0 & 1.0 & 1.0 & 1.0 & 1.0 & 1.0 & 1.0 & 1.0 & 1.0 & 1.0 & 1.0 & 1.0 & 1.0\\
\RotationArg{0.5}{(0.5, 0.5)} & 0.0 & 0.0 & 0.0 & 0.0 & 0.0 & 0.0 & 0.0 & 0.0 & 0.0 & 0.0 & 0.0 & 0.0 & 0.0 & 0.0 & 1.0 & 0.0 & 1.0 & 0.0 & 0.0\\
\RotationArg{10}{(0.5, 0.5)} & 1.0 & 1.0 & 1.0 & 1.0 & 1.0 & 1.0 & 1.0 & 1.0 & 1.0 & 1.0 & 1.0 & 1.0 & 1.0 & 1.0 & 1.0 & 1.0 & 1.0 & 1.0 & 1.0\\
\StretchArg{1.25}{1} & 0.0 & 0.004807692307692308 & 0.0 & 0.0 & 0.0 & 0.0 & 0.0 & 0.0 & 0.0 & 0.009615384615384616 & 0.0 & 0.009615384615384616 & 0.009615384615384616 & 0.0 & 0.004807692307692308 & 0.0 & 1.0 & 0.9326923076923077 & 0.5913461538461539\\
\StretchArg{1}{0.6} & 0.0 & 0.00009850275807722617 & 0.0 & 0.000049251379038613084 & 0.0 & 0.0 & 0.000049251379038613084 & 0.0 & 0.0 & 0.00009850275807722617 & 0.002364066193853428 & 0.001182033096926714 & 0.009062253743104806 & 0.0 & 0.0 & 0.0 & 0.009554767533490938 & 1.0 & 0.0076832151300236405\\
\StretchArg{1}{0.8} & 0.0 & 0.0058823529411764705 & 0.0 & 0.0 & 0.0 & 0.0 & 0.0 & 0.0 & 0.0 & 0.011764705882352941 & 0.0 & 0.011764705882352941 & 0.0058823529411764705 & 0.0 & 0.0 & 0.0 & 0.7235294117647059 & 0.9176470588235294 & 1.0\\
\end{tblr}%
}   
\caption{\PHOENIX}
\end{subfigure}

\vspace{1.5em}

\begin{subfigure}{\columnwidth}
\resizebox{\columnwidth}{!}{%
\begin{tblr}{colspec={lccccccccccccccccccc}, process=\funcBgColorDouble, row{2-20}={1.5em}, column{2-20} ={1.5em}, row{1}={cmd={\rotatebox{30}}}}
& \Identity & \Greyscale & \BilateralArg{10}{3} & \BilateralArg{125}{5} & \BilateralArg{80}{7} & \BrightnessArg{20}{0.8} & \GammaArg{0.5} & \Masked{bg}{\ColourwheelArg{90}} & \Masked{hair}{\ColourwheelArg{90}} & \MirrorArg{h}{} & \MotionArg{11}{0} & \MotionArg{11}{100} & \ResolutionArg{0.2} & \ResolutionArg{0.7} & \RotationArg{0.5}{(0.5, 0.5)} & \RotationArg{10}{(0.5, 0.5)} & \StretchArg{1.25}{1} & \StretchArg{1}{0.6} & \StretchArg{1}{0.8}\\
\Identity & 1.0 & 1.0 & 1.0 & 1.0 & 1.0 & 1.0 & 1.0 & 1.0 & 1.0 & 1.0 & 1.0 & 1.0 & 1.0 & 1.0 & 1.0 & 1.0 & 1.0 & 1.0 & 1.0 \\
\Greyscale & 0.0 & 1.0 & 0.0529595015576324 &  0.13395638629283488 &  0.16199376947040497 & 0.14953271028037382 & 0.22118380062305296 & 0.2803738317757009 & 0.18691588785046728 & 0.2803738317757009 & 0.3862928348909657 & 0.3862928348909657 & 0.34579439252336447 & 0.102803738317757 & 0.7289719626168224 & 0.9875389408099688 & 0.3613707165109034 & 0.67601246105919 & 0.3426791277258567 \\
\BilateralArg{10}{3} & 0.0 & 0.7727272727272727 & 1.0 & 0.8181818181818182 & 0.8181818181818182 & 0.6818181818181818 & 0.7727272727272727 & 0.7272727272727273 & 0.7272727272727273 & 0.9545454545454546 & 0.8636363636363636 & 0.8181818181818182 & 0.8181818181818182 & 0.8181818181818182 & 0.9090909090909091 & 0.9545454545454546 & 0.7727272727272727 & 0.8636363636363636 & 0.8181818181818182 \\
\BilateralArg{125}{5} & 0.0 & 0.6935483870967742 & 0.2903225806451613 & 1.0 & 0.9193548387096774 & 0.5 & 0.6612903225806451 & 0.6290322580645161 & 0.46774193548387094 & 0.8064516129032258 & 0.8548387096774194 & 0.8387096774193549 & 0.8709677419354839 & 0.6451612903225806 & 0.9354838709677419 & 0.9838709677419355 & 0.7419354838709677 & 0.7903225806451613 & 0.7419354838709677 \\
\BilateralArg{80}{7} & 0.0 & 0.7027027027027027 & 0.24324324324324326 & 0.7702702702702703 & 1.0 & 0.44594594594594594 & 0.6351351351351351 & 0.581081081081081 & 0.4594594594594595 & 0.7297297297297297 & 0.8513513513513513 & 0.8243243243243243 & 0.8783783783783784 & 0.5405405405405406 & 0.918918918918919 & 0.9594594594594594 & 0.7567567567567568 & 0.8108108108108109 & 0.7432432432432432 \\
\BrightnessArg{20}{0.8} & 0.0 & 0.7619047619047619 & 0.23809523809523808 & 0.49206349206349204 & 0.5238095238095238 & 1.0 & 0.7301587301587301 & 0.6031746031746031 & 0.5396825396825397 & 0.7142857142857143 & 0.746031746031746 & 0.6984126984126984 & 0.7936507936507936 & 0.3968253968253968 & 0.9523809523809523 & 0.9682539682539683 & 0.6984126984126984 & 0.8095238095238095 & 0.6984126984126984 \\
\GammaArg{0.5} & 0.0 & 0.6698113207547169 & 0.16037735849056603 & 0.3867924528301887 & 0.44339622641509435 & 0.4339622641509434 & 1.0 & 0.5188679245283019 & 0.37735849056603776 & 0.6226415094339622 & 0.6886792452830188 & 0.7358490566037735 & 0.7358490566037735 & 0.3490566037735849 & 0.8679245283018868 & 0.9716981132075472 & 0.6509433962264151 & 0.7547169811320755 & 0.6037735849056604 \\
\Masked{bg}{\ColourwheelArg{90}} & 0.0 & 0.7894736842105263 & 0.14035087719298245 & 0.34210526315789475 & 0.37719298245614036 & 0.3333333333333333 & 0.4824561403508772 & 1.0 & 0.3508771929824561 & 0.5526315789473685 & 0.6754385964912281 & 0.6403508771929824 & 0.6578947368421053 & 0.2982456140350877 & 0.8947368421052632 & 0.9649122807017544 & 0.6052631578947368 & 0.8333333333333334 & 0.543859649122807 \\
\Masked{hair}{\ColourwheelArg{90}} & 0.0 & 0.8450704225352113 & 0.22535211267605634 & 0.4084507042253521 & 0.4788732394366197 & 0.4788732394366197 & 0.5633802816901409 & 0.5633802816901409 & 1.0 & 0.6338028169014085 & 0.6901408450704225 & 0.676056338028169 & 0.7605633802816901 & 0.39436619718309857 & 0.8450704225352113 & 0.9859154929577465 & 0.6901408450704225 & 0.8028169014084507 & 0.6197183098591549 \\
\MirrorArg{h}{} & 0.0 & 0.6976744186046512 & 0.16279069767441862 & 0.3875968992248062 & 0.4186046511627907 & 0.3488372093023256 & 0.5116279069767442 & 0.4883720930232558 & 0.3488372093023256 & 1.0 & 0.6744186046511628 & 0.6589147286821705 & 0.6589147286821705 & 0.31007751937984496 & 0.8837209302325582 & 0.9844961240310077 & 0.6356589147286822 & 0.7906976744186046 & 0.6356589147286822 \\
\MotionArg{11}{0} & 0.0 & 0.6492146596858639 & 0.09947643979057591 & 0.2774869109947644 & 0.3298429319371728 & 0.24607329842931938 & 0.38219895287958117 & 0.4031413612565445 & 0.25654450261780104 & 0.45549738219895286 & 1.0 & 0.6178010471204188 & 0.680628272251309 & 0.21465968586387435 & 0.837696335078534 & 0.9790575916230366 & 0.4973821989528796 & 0.7958115183246073 & 0.4869109947643979 \\
\MotionArg{11}{100} & 0.0 & 0.7005649717514124 & 0.1016949152542373 & 0.2937853107344633 & 0.3446327683615819 & 0.24858757062146894 & 0.4406779661016949 & 0.4124293785310734 & 0.2711864406779661 & 0.480225988700565 & 0.6666666666666666 & 1.0 & 0.7062146892655368 & 0.24293785310734464 & 0.8757062146892656 & 0.9774011299435028 & 0.5536723163841808 & 0.8135593220338984 & 0.5254237288135594 \\
\ResolutionArg{0.2} & 0.0 & 0.6201117318435754 & 0.1005586592178771 & 0.3016759776536313 & 0.36312849162011174 & 0.27932960893854747 & 0.43575418994413406 & 0.41899441340782123 & 0.3016759776536313 & 0.4748603351955307 & 0.7262569832402235 & 0.6983240223463687 & 1.0 & 0.24581005586592178 & 0.8659217877094972 & 0.9720670391061452 & 0.5363128491620112 & 0.770949720670391 & 0.5307262569832403 \\
\ResolutionArg{0.7} & 0.0 & 0.6470588235294118 & 0.35294117647058826 & 0.7843137254901961 & 0.7843137254901961 & 0.49019607843137253 & 0.7254901960784313 & 0.6666666666666666 & 0.5490196078431373 & 0.7843137254901961 & 0.803921568627451 & 0.8431372549019608 & 0.8627450980392157 & 1.0 & 0.9215686274509803 & 0.9803921568627451 & 0.7450980392156863 & 0.7843137254901961 & 0.6862745098039216 \\
\RotationArg{0.5}{(0.5, 0.5)} & 0.0 & 0.47657841140529533 & 0.04073319755600815 & 0.11812627291242363 & 0.1384928716904277 & 0.12219959266802444 & 0.18737270875763748 & 0.20773930753564154 & 0.12219959266802444 & 0.23217922606924643 & 0.3258655804480652 & 0.31568228105906315 & 0.31568228105906315 & 0.09572301425661914 & 1.0 & 0.9938900203665988 & 0.2790224032586558 & 0.6456211812627292 & 0.2606924643584521 \\
\RotationArg{10}{(0.5, 0.5)} & 0.0 & 0.38658536585365855 & 0.025609756097560974 & 0.07439024390243902 & 0.08658536585365853 & 0.07439024390243902 & 0.12560975609756098 & 0.13414634146341464 & 0.08536585365853659 & 0.1548780487804878 & 0.2280487804878049 & 0.21097560975609755 & 0.2121951219512195 & 0.06097560975609756 & 0.5951219512195122 & 1.0 & 0.2121951219512195 & 0.5829268292682926 & 0.19878048780487806 \\
\StretchArg{1.25}{1} & 0.0 & 0.6590909090909091 & 0.09659090909090909 & 0.26136363636363635 & 0.3181818181818182 & 0.25 & 0.39204545454545453 & 0.39204545454545453 & 0.2784090909090909 & 0.4659090909090909 & 0.5397727272727273 & 0.5568181818181818 & 0.5454545454545454 & 0.2159090909090909 & 0.7784090909090909 & 0.9886363636363636 & 1.0 & 0.8863636363636364 & 0.8125 \\
\StretchArg{1}{0.6} & 0.0 & 0.45208333333333334 & 0.03958333333333333 & 0.10208333333333333 & 0.125 & 0.10625 & 0.16666666666666666 & 0.19791666666666666 & 0.11875 & 0.2125 & 0.31666666666666665 & 0.3 & 0.2875 & 0.08333333333333333 & 0.6604166666666667 & 0.9958333333333333 & 0.325 & 1.0 & 0.30625 \\
\StretchArg{1}{0.8} & 0.0 & 0.6748466257668712 & 0.11042944785276074 & 0.2822085889570552 & 0.3374233128834356 & 0.26993865030674846 & 0.39263803680981596 & 0.3803680981595092 & 0.26993865030674846 & 0.5030674846625767 & 0.5705521472392638 & 0.5705521472392638 & 0.5828220858895705 & 0.2147239263803681 & 0.7852760736196319 & 1.0 & 0.8773006134969326 & 0.901840490797546 & 1.0 \\
\end{tblr}%
}
\caption{\FLIC}
\end{subfigure}

\caption{\SubRate{\metRule_1}{\metRule_2}, with $\metRule_1$ left and $\metRule_2$ top, for each dataset, using body pose landmarks and a 0.2 error threshold.}
\label{fig:rq3_subsumption}
\end{figure}

To analyse the contribution of different metamorphic relations when using \proposed we can look at their \textit{subsumption rates}. A rule $\metRule_1$ subsumes a rule $\metRule_2$ if, when $\metRule_2$ detects a failure for an image then $\metRule_1$ also detected that failure. This means that running $\metRule_1$ is sufficient to detect the failures $\metRule_2$ detects. Given two rules, $\metRule_1$ and $\metRule_2$, we explore the degree to which $\metRule_1$ subsumes $\metRule_2$ using the subsumption rate \SubRate{\metRule_1}{\metRule_2}:
\[
\SubRate{\metRule_1}{\metRule_2}=
\begin{cases}
\text{if $\metRule_1$ is never violated,} 1\\
\text{else, }\frac{\text{\# images violating $\metRule_1$ and $\metRule_2$}}{\text{\# images violating $\metRule_1$}}
\end{cases}
\]

Figure~\ref{fig:rq3_subsumption} shows the subsumption rates of all pairs of rules in \exampleRels for each dataset for a set value of the error threshold (0.2). These results show that the subsumption rates between the rules are very dependent on the dataset used, indicating that the different rules make different contributions to \proposed, i.e., they find different types of errors. 

Figure~\ref{fig:rq3_motion_blur_subsumptions} focuses on the different settings of the \Motion rule and how they subsume each other. While one might expect that settings that introduce more discrete blur would clearly subsume settings that introduce stronger blur, the results do not show such a clear result. This is probably due to the complex nature of \mediapipe\holistic, which means that the direct effects of a transformation on the system's output are not easily predictable -- further illustrating the value of the proposed testing procedure.
\begin{figure}[!t]
\begin{subfigure}[t]{.5\columnwidth}
\centering
\resizebox{\textwidth}{!}{%
\begin{tblr}{colspec={lccccccccccccccc}, process=\funcBgColorDouble, row{2-16}={1.5em}, column{2-16} ={1.5em}, row{1}={cmd={\rotatebox{45}}}}
& \MotionArg{5}{0} & \MotionArg{5}{40} & \MotionArg{5}{70} & \MotionArg{5}{100} & \MotionArg{7}{0} & \MotionArg{7}{40} & \MotionArg{7}{70} & \MotionArg{7}{100} & \MotionArg{9}{0} &  \MotionArg{9}{40} &  \MotionArg{9}{70} &  \MotionArg{9}{100} & \MotionArg{11}{0} & \MotionArg{11}{70} & \MotionArg{11}{100} \\ 
\MotionArg{5}{0} &  1.0 & 0.0 & 0.0 & 0.0 & 0.0 & 0.0 & 0.0 & 0.0 & 0.0 & 0.0 & 0.0 & 0.0 & 0.0 & 0.0 & 0.0 \\
\MotionArg{5}{40} &  1.0 & 1.0 & 1.0 & 1.0 & 1.0 & 1.0 & 1.0 & 1.0 & 1.0 & 1.0 & 1.0 & 1.0 & 1.0 & 1.0 & 1.0 \\
\MotionArg{5}{70} &  1.0 & 1.0 & 1.0 & 1.0 & 1.0 & 1.0 & 1.0 & 1.0 & 1.0 & 1.0 & 1.0 & 1.0 & 1.0 & 1.0 & 1.0 \\
\MotionArg{5}{100} &  0.0 & 0.0 & 0.0 & 1.0 & 0.0 & 0.0 & 0.0 & 0.5 & 0.0 & 0.0 & 0.0 & 0.0 & 0.0 & 0.0 & 0.0 \\
\MotionArg{7}{0} &  0.0 & 0.0 & 0.0 & 0.0 & 1.0 & 0.6666666666666666 & 0.0 & 0.0 & 1.0 & 1.0 & 0.6666666666666666 & 0.0 & 1.0 & 0.6666666666666666 & 0.3333333333333333 \\
\MotionArg{7}{40} &  0.0 & 0.0 & 0.0 & 0.0 & 0.4 & 1.0 & 0.0 & 0.0 & 0.6 & 1.0 & 0.6 & 0.0 & 0.8 & 0.6 & 0.4 \\
\MotionArg{7}{70} &  0.0 & 0.0 & 0.0 & 0.0 & 0.0 & 0.0 & 1.0 & 1.0 & 0.0 & 1.0 & 1.0 & 1.0 & 0.0 & 0.0 & 1.0 \\
\MotionArg{7}{100} &  0.0 & 0.0 & 0.0 & 0.5 & 0.0 & 0.0 & 0.5 & 1.0 & 0.0 & 0.5 & 0.5 & 0.5 & 0.0 & 0.0 & 0.5 \\
\MotionArg{9}{0} &  0.0 & 0.0 & 0.0 & 0.0 & 0.25 & 0.25 & 0.0 & 0.0 & 1.0 & 0.5833333333333334 & 0.25 & 0.0 & 1.0 & 0.3333333333333333 & 0.16666666666666666 \\
\MotionArg{9}{40} &  0.0 & 0.0 & 0.0 & 0.0 & 0.16666666666666666 & 0.2777777777777778 & 0.05555555555555555 & 0.05555555555555555 & 0.3888888888888889 & 1.0 & 0.3333333333333333 & 0.05555555555555555 & 0.5555555555555556 & 0.4444444444444444 & 0.2777777777777778 \\
\MotionArg{9}{70} &  0.0 & 0.0 & 0.0 & 0.0 & 0.2857142857142857 & 0.42857142857142855 & 0.14285714285714285 & 0.14285714285714285 & 0.42857142857142855 & 0.8571428571428571 & 1.0 & 0.14285714285714285 & 0.42857142857142855 & 0.8571428571428571 & 0.5714285714285714 \\
\MotionArg{9}{100} &  0.0 & 0.0 & 0.0 & 0.0 & 0.0 & 0.0 & 0.3333333333333333 & 0.3333333333333333 & 0.0 & 0.3333333333333333 & 0.3333333333333333 & 1.0 & 0.0 & 0.3333333333333333 & 1.0 \\
\MotionArg{11}{0} &  0.0 & 0.0 & 0.0 & 0.0 & 0.03896103896103896 & 0.05194805194805195 & 0.0 & 0.0 & 0.15584415584415584 & 0.12987012987012986 & 0.03896103896103896 & 0.0 & 1.0 & 0.07792207792207792 & 0.025974025974025976 \\
\MotionArg{11}{40} &  0.0 & 0.0 & 0.0 & 0.0 & 0.041666666666666664 & 0.0625 & 0.0 & 0.0 & 0.08333333333333333 & 0.16666666666666666 & 0.125 & 0.020833333333333332 & 0.125 & 1.0 & 0.125 \\
\MotionArg{11}{100} &  0.0 & 0.0 & 0.0 & 0.0 & 0.030303030303030304 & 0.06060606060606061 & 0.030303030303030304 & 0.030303030303030304 & 0.06060606060606061 & 0.15151515151515152 & 0.12121212121212122 & 0.09090909090909091 & 0.06060606060606061 & 0.18181818181818182 & 1.0 \\
\end{tblr}%
}
\caption{\PHOENIX}
\label{fig:motion_blur_subsumption_phoenix}
\end{subfigure}%
\begin{subfigure}[t]{.5\columnwidth}
\centering
\resizebox{\textwidth}{!}{%
\begin{tblr}{colspec={lccccccccccccccc}, process=\funcBgColorDouble, row{2-16}={1.5em}, column{2-16} ={1.5em}, row{1}={cmd={\rotatebox{30}}}}
& \MotionArg{5}{0} & \MotionArg{5}{40} & \MotionArg{5}{70} & \MotionArg{5}{100} & \MotionArg{7}{0} & \MotionArg{7}{40} & \MotionArg{7}{70} & \MotionArg{7}{100} & \MotionArg{9}{0} &  \MotionArg{9}{40} &  \MotionArg{9}{70} &  \MotionArg{9}{100} & \MotionArg{11}{0} & \MotionArg{11}{70} & \MotionArg{11}{100}\\ 
\MotionArg{5}{0} & 1.0 & 0.863013698630137 & 0.6712328767123288 & 0.6575342465753424 & 0.8904109589041096 & 0.9041095890410958 & 0.7808219178082192 & 0.7123287671232876 & 0.9041095890410958 & 0.8767123287671232 & 0.7671232876712328 & 0.7397260273972602 & 0.9178082191780822 & 0.7945205479452054 & 0.8082191780821918 \\
\MotionArg{5}{40} & 0.7590361445783133 & 1.0 & 0.7108433734939759 & 0.6385542168674698 & 0.7951807228915663 & 0.927710843373494 & 0.7951807228915663 & 0.6987951807228916 & 0.8313253012048193 & 0.9036144578313253 & 0.8192771084337349 & 0.7469879518072289 & 0.8433734939759037 & 0.8313253012048193 & 0.8072289156626506 \\
\MotionArg{5}{70} & 0.6533333333333333 & 0.7866666666666666 & 1.0 & 0.8 & 0.7733333333333333 & 0.8266666666666667 & 0.8666666666666667 & 0.84 & 0.84 & 0.8933333333333333 & 0.84 & 0.84 & 0.8266666666666667 & 0.88 & 0.8533333333333334 \\
\MotionArg{5}{100} & 0.6666666666666666 & 0.7361111111111112 & 0.8333333333333334 & 1.0 & 0.7777777777777778 & 0.7777777777777778 & 0.8611111111111112 & 0.875 & 0.7916666666666666 & 0.8472222222222222 & 0.8333333333333334 & 0.8888888888888888 & 0.8333333333333334 & 0.875 & 0.875 \\
\MotionArg{7}{0} & 0.65 & 0.66 & 0.58 & 0.56 & 1.0 & 0.73 & 0.68 & 0.64 & 0.94 & 0.8 & 0.72 & 0.68 & 0.9 & 0.78 & 0.79 \\
\MotionArg{7}{40} & 0.6666666666666666 & 0.7777777777777778 & 0.6262626262626263 & 0.5656565656565656 & 0.7373737373737373 & 1.0 & 0.7474747474747475 & 0.6363636363636364 & 0.8383838383838383 & 0.9090909090909091 & 0.7777777777777778 & 0.7171717171717171 & 0.8686868686868687 & 0.8282828282828283 & 0.8181818181818182 \\
\MotionArg{7}{70} & 0.59375 & 0.6875 & 0.6770833333333334 & 0.6458333333333334 & 0.7083333333333334 & 0.7708333333333334 & 1.0 & 0.8229166666666666 & 0.84375 & 0.8645833333333334 & 0.96875 & 0.875 & 0.8645833333333334 & 0.9479166666666666 & 0.9375 \\
\MotionArg{7}{100} & 0.5531914893617021 & 0.6170212765957447 & 0.6702127659574468 & 0.6702127659574468 & 0.6808510638297872 & 0.6702127659574468 & 0.8404255319148937 & 1.0 & 0.8085106382978723 & 0.7659574468085106 & 0.8617021276595744 & 0.8936170212765957 & 0.8085106382978723 & 0.851063829787234 & 0.9042553191489362 \\
\MotionArg{9}{0} & 0.4370860927152318 & 0.45695364238410596 & 0.41721854304635764 & 0.37748344370860926 & 0.6225165562913907 & 0.5496688741721855 & 0.5364238410596026 & 0.5033112582781457 & 1.0 & 0.6556291390728477 & 0.5960264900662252 & 0.5827814569536424 & 0.9403973509933775 & 0.6754966887417219 & 0.7152317880794702 \\
\MotionArg{9}{40} & 0.5245901639344263 & 0.6147540983606558 & 0.5491803278688525 & 0.5 & 0.6557377049180327 & 0.7377049180327869 & 0.680327868852459 & 0.5901639344262295 & 0.8114754098360656 & 1.0 & 0.7622950819672131 & 0.7049180327868853 & 0.8442622950819673 & 0.8442622950819673 & 0.819672131147541 \\
\MotionArg{9}{70} & 0.45161290322580644 & 0.5483870967741935 & 0.5080645161290323 & 0.4838709677419355 & 0.5806451612903226 & 0.6209677419354839 & 0.75 & 0.6532258064516129 & 0.7258064516129032 & 0.75 & 1.0 & 0.7903225806451613 & 0.7741935483870968 & 0.9274193548387096 & 0.9032258064516129 \\
\MotionArg{9}{100} & 0.4426229508196721 & 0.5081967213114754 & 0.5163934426229508 & 0.5245901639344263 & 0.5573770491803278 & 0.5819672131147541 & 0.6885245901639344 & 0.6885245901639344 & 0.7213114754098361 & 0.7049180327868853 & 0.8032786885245902 & 1.0 & 0.7459016393442623 & 0.8688524590163934 & 0.9590163934426229 \\
\MotionArg{11}{0} & 0.3507853403141361 & 0.36649214659685864 & 0.32460732984293195 & 0.31413612565445026 & 0.4712041884816754 & 0.450261780104712 & 0.43455497382198954 & 0.39790575916230364 & 0.743455497382199 & 0.5392670157068062 & 0.5026178010471204 & 0.47643979057591623 & 1.0 & 0.612565445026178 & 0.6178010471204188 \\
\MotionArg{11}{70} & 0.3493975903614458 & 0.41566265060240964 & 0.39759036144578314 & 0.3795180722891566 & 0.46987951807228917 & 0.4939759036144578 & 0.5481927710843374 & 0.4819277108433735 & 0.6144578313253012 & 0.6204819277108434 & 0.6927710843373494 & 0.6385542168674698 & 0.7048192771084337 & 1.0 & 0.8313253012048193 \\
\MotionArg{11}{100} & 0.3333333333333333 & 0.3785310734463277 & 0.3615819209039548 & 0.3559322033898305 & 0.4463276836158192 & 0.4576271186440678 & 0.5084745762711864 & 0.480225988700565 & 0.6101694915254238 & 0.5649717514124294 & 0.632768361581921 & 0.6610169491525424 & 0.6666666666666666 & 0.7796610169491526 & 1.0 \\
\end{tblr}%
}
\caption{\FLIC}
\label{fig:motion_blur_subsumption_flic}
\end{subfigure}%
\caption{Subsumption rates of various motion blur settings}
\label{fig:rq3_motion_blur_subsumptions}
\end{figure}

The data in Table~\ref{tab:how_many_failed_relations_both} confirms that the different rules reveal different types of faults. Indeed, when using a reasonable error threshold (not so low that nearly all images lead to rule violations), most images lead to a violation of only a small number of rules. This again suggests that the different rules assess different aspects of the system as, in most cases, the same input leads to a violation only for specific rules.
\begin{table*}[tbp]
    \centering
    \caption{Images that violate different numbers of rules at given error thresholds in \PHOENIX (\PHOENIXShort) and \FLIC (\FLICShort)}
    \label{tab:how_many_failed_relations_both}
{\setlength{\tabcolsep}{3pt}\scriptsize
\begin{tabular}{@{}lrrrrrrrrrrrrrrrrrrrrrr@{}} 
\toprule 
\multirow{2}{*}{\diagbox[width=2.6cm, height=3.2em]{\textbf{\# failed rules}}{\textbf{Error}\\\textbf{thresh.}}}
& \multicolumn{2}{c}{0.01} & \multicolumn{2}{c}{0.05} & \multicolumn{2}{c}{0.1} & \multicolumn{2}{c}{0.2} & \multicolumn{2}{c}{0.3} & \multicolumn{2}{c}{0.7} & \multicolumn{2}{c}{1.0} & \multicolumn{2}{c}{2.0} & \multicolumn{2}{c}{5.0} & \multicolumn{2}{c}{20.0} & \multicolumn{2}{c}{$\inf$} \\
\cmidrule(r){2-3}\cmidrule(rl){4-5}\cmidrule(rl){6-7}\cmidrule(rl){8-9}\cmidrule(rl){10-11}\cmidrule(rl){12-13}\cmidrule(rl){14-15}\cmidrule(rl){16-17}\cmidrule(rl){18-19}\cmidrule(rl){20-21}\cmidrule(rl){22-23}
& \FLICShort & \PHOENIXShort & \FLICShort & \PHOENIXShort & \FLICShort & \PHOENIXShort & \FLICShort & \PHOENIXShort & \FLICShort & \PHOENIXShort & \FLICShort & \PHOENIXShort & \FLICShort & \PHOENIXShort & \FLICShort & \PHOENIXShort & \FLICShort & \PHOENIXShort & \FLICShort & \PHOENIXShort & \FLICShort & \PHOENIXShort\\\midrule
0 & 2 & 0 & 2 & 0 & 2 & 0 & 2 & 0 & 2 & 0 & 9 & 11996 & 56 & 17794 & 98 & 18767 & 135 & 18813 & 138 & 18821 & 138 &  18821 \\
1 & 0 & 0 & 0 & 0 & 0 & 0 & 0 & 0 & 0 & 0 & 35 & 23192 & 51 & 28258 & 77 & 28789 & 113 & 28759 & 121 & 28752 & 122 &  28752 \\
2 & 0 & 0 & 0 & 0 & 0 & 1 & 0 & 10363 & 0 & 14486 & 52 & 11742 & 54 & 7144 & 44 & 6180 & 74 & 6171 & 84 & 6171 & 84 &  6171 \\
3 & 1 & 0 & 1 & 0 & 1 & 1401 & 1 & 18521 & 1 & 24116 & 48 & 5985 & 52 & 1616 & 47 & 1334 & 44 & 1330 & 47 & 1329 & 49 &  1329 \\
4 & 0 & 0 & 0 & 0 & 0 & 5129 & 0 & 15246 & 15 & 11806 & 48 & 1956 & 36 & 617 & 33 & 446 & 41 & 443 & 46 & 443 & 45 &  443 \\ 
5 to 30 (25\%)  & 9 & 0 & 198 & 53427 & 534 & 49229 & 635 & 11644 & 643 & 5367 & 494 & 904 & 446 & 346 & 414 & 259 & 354 & 259 & 355 & 259 & 353 &  259 \\
31 to 61 (50\%)  & 16 & 0 & 373 & 2668 & 206 & 16 & 136 & 1 & 118 & 0 & 111 & 0 & 104 & 0 & 92 & 0 & 59 & 0 & 36 & 0 & 36 &  0 \\
62 to 91 (75\%)  & 5 & 2635 & 163 & 38 & 58 & 0 & 46 & 0 & 41 & 0 & 35 & 0 & 30 & 0 & 26 & 0 & 14 & 0 & 7 & 0 & 7 &  0 \\
92 to 121  & 464 & 53684 & 94 & 2 & 46 & 0 & 32 & 0 & 25 & 0 & 12 & 0 & 11 & 0 & 10 & 0 & 6 & 0 & 2 & 0 & 2 &  0 \\
122 (100\%)  & 338 & 42 & 27 & 0 & 7 & 0 & 1 & 0 & 0 & 0 & 0 & 0 & 0 & 0 & 0 & 0 & 0 & 0 & 0 & 0 & 0 &  0 \\ 
\bottomrule 
\end{tabular}}
\end{table*}

\section{Related Work}
\label{sec:rw}

The use of machine learning systems is becoming increasingly common in many diverse fields. Testing strategies for these systems are therefore being more widely recognised as a necessity, especially in safety critical systems~\cite{kassab2018testing}. In particular, these systems are tested with different focuses in mind such as: robustness to degradations \cite{hand-pose}\cite{hendrycks2019benchmarkingneuralnetworkrobustness}\cite{semantic_perturbations}\cite{object-centric-learning}, testing for faults when using ML frameworks~\cite{humbatova2020taxonomy}, testing the adequacy of the datasets used to train and test these systems~\cite{kim2023evaluating} and testing the biases embedded in these systems~\cite{birhane2022auditing, buolamwini2018gender}.

In particular, metamorphic testing is a convenient method to circumvent the oracle problem found when attempting to test many of these systems. This method has been used to test a wide range of ML systems, including: autonomous delivery robots~\cite{laurent2024metamorphic}, image classification systems~\cite{naidu2021metamorphic}, recommender systems~\cite{recommended-systems-testing}, search engines~\cite{testingSearchEngines} and chat bots~\cite{chatbotTesting}.

More particularly, when looking at pose estimation systems, we can find previous work focused on the pose estimation of hands \cite{hand-pose}, and on predicting the trajectory of humans, done by autonomous systems \cite{human-trajectory-metTest}.

When looking at the particular model tested in this work, it has previously been explicitly evaluated on individuals from different regions of the world which has been reported in the model's corresponding Model Card~\cite{modelcard2021mediapipe, mitchell2019model}. 

One of the main differences between the related work and the framework presented here is that we avoid the need for ground truth labelled data, meaning that \proposed can work with datasets without ground truth, which has not been done in previous metamorphic testing applied to pose estimation. We also do not focus only on robustness to degradations. We present a framework that can be used for various specific interests of the applications, which can be image degradation, but can encompass a wide range of potential concerns from fairness issues to sensitivity to image quality. Even if we do not focus on suggesting novel metamorphic rules, when novel metamorphic rules are needed for a specific application, \proposed can be applied to new datasets without the expense of labelling ground truth keypoints.

\section{Threats to validity}
\label{sec:threats}
We list (and classify following\cite{Feldt2010ValidityTI}) threats to the validity of our work here:

\begin{itemize}
    \item \textit{Construct Validity:} A possible threat is the validity of the metrics used to reach our conclusions. The results would be different if instead of counting the number of inputs that fail with at least one metamorphic rule, we required multiple failures for each input to be counted. They would also look different depending on the metamorphic relations we take into account, or if the modified images are out of scope of the system under test. The image aggregation metric can also change the results depending on whether we measure the mean, median, minimum or maximum error of all keypoints in an output. 
    
    To mitigate this, we consider the median error as an image aggregation metric, which is less sensitive to outliers compared to the mean or the maximum. We also show the number of failed inputs for different selections of metamorphic relations. We choose to count the inputs when they fail for only one metamorphic relation instead of many because this is enough to show that the framework found a problem that would be relevant to the application engineers. Finally, we show how a wide and varied selection of different metamorphic relations do not completely subsume each other, so the errors found come from the test with various different rules, and not only one rule finding all the problems.

    We also note that in this work we are not trying to find specific problems in the SUT, and so the influence of the metrics discussed here is reduced. They show that \proposed can be used to find problems in such systems, but the focus is not on which specific problems we found.
        
    \item \textit{Internal Validity:} Another threat is that the results shown were found by chance or due to mistakes in our implementation, and not because of the causes that we expect.

    To mitigate this, we have carefully inspected our implementation, and manually confirmed each step done until the results aggregation. We have also added an \textit{identity} metamorphic relation as a sanity check to ensure that the system results do not change when we run the system with the same input twice, in which case we would need to run our experiments a number of times to reach a statistically significant conclusion.        
        
    \item \textit{External Validity:} The current work may not generalise to other datasets or systems beyond the ones in this work. It is also not explored how effective it would be when used in a real application setting with a specific set of relevant input features and output requirements. In general, \texttt{DL} models are evolving rapidly and so it is common that many methods are rendered obsolete quickly.

    To mitigate these, we have designed this framework in a way that it is highly dataset and system agnostic: any image is processed equally, not depending on its colour, number or type of subjects, etc.; and \proposed tests the SUT as a black-box. The only change required to use different ground truth annotations, or different systems, is due to the non-standardised pose estimation outputs and annotations, which is unavoidable. We have also left not only the specific metamorphic relations but also the error metric open to modification for different applications, facilitating adaptability to different contexts.
\end{itemize}

\section{Conclusions and Future Work}
\label{sec:concAndFuture}

In this work we propose \proposed, a metamorphic testing-based framework for pose estimation systems. We have focused on making this framework both system- and dataset-agnostic, due to the rapid developments of the area of pose estimation systems. This should help the approach remain relevant even if architectures or methodological changes occur in the area. 

We applied \proposed to \mediapipe\holistic using two datasets from different application domains. Results show that \proposed can detect failures of the system without ground truth data, i.e., without human annotations. They also show that, by leaving the definition of the metamorphic rules and the metrics used to assess violation of these rules open to the user, \proposed is adaptable and can test pose estimation systems for different use cases.

As future work, we plan to apply \proposed to different systems and use cases. This will first confirm the framework's adaptability. We also plan to show that, by using application-specific rules, practitioners who build these pose estimation systems can gain a better understanding of how they operate and even use this information towards repairing their systems.

\bibliographystyle{IEEEtran}
\bibliography{references}
\balance
\end{document}

%% file: heatMapPreap.tex
\usepackage{pgf}
\usepackage{multicol}
\usepackage{colortbl}
\usepackage{hhline}
\usepackage{tabularray}
\usepackage{rotating}
\UseTblrLibrary{functional}
\usepackage{tikz}

\newcommand{\whitebox}[1]{\ifthenelse{\equal{#1}{-}}{#1}{\fcolorbox{black}{white}{#1}}}

\IgnoreSpacesOn
\prgNewFunction \funcBgColor {} {
  \intStepOneInline {1} {\arabic{rowcount}} {
    \intStepOneInline {1} {\arabic{colcount}} {
        \regexMatchT {^-?\d+(?:\.\d+)?\Z|^-?\d+\/\d+\Z} 
        {\cellGetText {##1} {####1}} {
            \fpSet \cellVal {\cellGetText {##1} {####1}}
            \cellSetText {##1} {####1} {\fpEval {round(\cellVal, 2)}}
            \fpSet \colorVal {100 * (\cellVal - \heatMapMin) / (\heatMapMax - \heatMapMin)}
            \tlSet \colorValtl {\fpUse \colorVal} 
            \pgfmathparse{\colorValtl}
            \cellSetStyle {##1} {####1} {bg=red3!\pgfmathresult!blue3}
        }
    }
  }
}
\prgNewFunction \funcBgColorAutoPercent {} {
    \fpZeroNew \maxVal
    \fpZeroNew \minVal
  \intStepOneInline {1} {\arabic{colcount}} {
    \intStepOneInline {1} {\arabic{rowcount}-1} {
        \regexMatchT {^-?\d+(?:\.\d+)?\Z|^-?\d+\/\d+\Z} {\cellGetText {##1} {####1}} {
            \fpSet \currVal {\cellGetText {##1} {####1}}
            \fpSet \maxVal {max(\maxVal, \currVal)}
            \fpSet \minVal {min(\minVal, \currVal)}
        }
    }
  }
  \intStepOneInline {1} {\arabic{rowcount}} {
    \intStepOneInline {1} {\arabic{colcount}} {
        \regexMatchT {^-?\d+(?:\.\d+)?\Z|-?\d+\/\d+} {\cellGetText {##1} {####1}} {
            \fpSet \cellVal {\cellGetText {##1} {####1}}
            \strSet \cellStr {\fpEval {round(\cellVal * 100, 2)}}
            \strConcat \lTmpaStr \cellStr \cPercentStr
            \cellSetText {##1} {####1} {\strUse \lTmpaStr}
            \fpSet \colorVal {max(0, min(100 * (\cellVal - \minVal) / (\maxVal - \minVal),100))}
            \tlSet \colorValtl {\fpUse \colorVal} 
            \pgfmathparse{\colorValtl}
            \cellSetStyle {##1} {####1} {bg=red4!\pgfmathresult!blue4, fg=white}
        }
    }
  }
}
\prgNewFunction \funcBgColorAuto {} {
    \fpZeroNew \maxVal
    \fpZeroNew \minVal
  \intStepOneInline {1} {\arabic{colcount}} {
    \intStepOneInline {1} {\arabic{rowcount}-1} {
        \regexMatchT {^-?\d+(?:\.\d+)?\Z|^-?\d+\/\d+\Z} {\cellGetText {##1} {####1}} {
            \fpSet \currVal {\cellGetText {##1} {####1}}
            \fpSet \maxVal {max(\maxVal, \currVal)}
            \fpSet \minVal {min(\minVal, \currVal)}
        }
    }
  }
  \intStepOneInline {1} {\arabic{rowcount}} {
    \intStepOneInline {1} {\arabic{colcount}} {
        \regexMatchT {^-?\d+(?:\.\d+)?\Z|-?\d+\/\d+} {\cellGetText {##1} {####1}} {
            \fpSet \cellVal {\cellGetText {##1} {####1}}
            \strSet \cellStr {\fpEval {round(\cellVal, 2)}}
            \cellSetText {##1} {####1} {\strUse \cellStr}
            \fpSet \colorVal {max(0, min(100 * (\cellVal - \minVal) / (\maxVal - \minVal),100))}
            \tlSet \colorValtl {\fpUse \colorVal} 
            \pgfmathparse{\colorValtl}
            \cellSetStyle {##1} {####1} {bg=red4!\pgfmathresult!blue4, fg=white}
        }
    }
  }
}
\prgNewFunction \funcBgColorDouble {} {
  \intStepOneInline {1} {\arabic{rowcount}} {
    \intStepOneInline {1} {\arabic{colcount}} {
        \regexMatchT {^-?\d+(?:\.\d+)?\Z|^-?\d+\/\d+\Z} {\cellGetText {##1} {####1}} {
            \fpSet \cellVal {\cellGetText {##1} {####1}}
            \cellSetText {##1} {####1} {\fpEval {round(\cellVal, 2)}}
            \fpCompareTF {\cellVal} > {\heatMapMax/2} {\fpSet \colorVal {100 * (\cellVal - \heatMapMax / 2) / (\heatMapMax / 2)}} {\fpSet \colorVal {100 * \cellVal / (\heatMapMax / 2)}}
            \tlSet \colorValtl {\fpUse \colorVal} 
            \pgfmathparse{\colorValtl}            
            \fpCompareTF {\cellVal} > {\heatMapMax/2}
            {\cellSetStyle {##1} {####1} {bg=yellow!\pgfmathresult!red!70}}
            {\cellSetStyle {##1} {####1} {bg=red!\pgfmathresult!blue!70}}
        }
    }
  }
}
\IgnoreSpacesOff

%% file: introduction.tex
\section{Introduction}
\label{sec:intro}
 \textit{Pose Estimation Systems} has applications in medicine~\cite{LITJENS201760}, sign language recognition~\cite{holmes2023scarcity} and high stakes sports events~\cite{martin2021automated},  requiring them to be well-tested in order to provide a substantiated assessment of the correct behaviour when applied to sensitive domains. Such systems need to perform correctly and as expected under a variety of conditions. This work aims to provide practitioners with a means of assessing this.

In this work we propose \proposed, a metamorphic testing framework to test pose estimation systems without the significant cost of manual data labelling. Additionally, we propose a non exhaustive set of metamorphic rules for \proposed, including flexible metrics to assess violations. These rules allow practitioners to apply various commonly encountered image changes and can easily be extended by users, should they want to explore different aspects of the system.

We apply \proposed to Mediapipe Hollistic~\cite{researchMediaPipeHolistic}, a widely used state-of-the-art pose estimation system, on datasets from the literature used to train and assess human pose estimation systems in different domains. We show that our proposed framework can find numerous faulty outputs from the system. Results show that \proposed provides results on par with classic, human annotated ground truth-based testing on the \FLIC dataset. Furthermore, we illustrate how analysis of the results can highlight elements of the input that impact the system's performance. Such analysis can then help practitioners better understand the settings under which the system functions properly, and can thus be used with confidence.

The remainder of this paper is organised as follows. First, Section~\ref{sec:background} provides an overview of the context and concepts that underpin this work. Next, Section~\ref{sec:contribution} describes our proposed metamorphic testing framework for pose estimation systems, and Section~\ref{sec:metrules} describes a non-exhaustive set of possible metamorphic relations for this system. Section~\ref{sec:exps} describes the experiments performed to evaluate the framework, and Section~\ref{sec:results} describes and analyses the results of these experiments. Section~\ref{sec:rw} gives an overview of related work, while Section~\ref{sec:threats} describes threats to the validity of this work, and steps taken to address them. Finally, Section~\ref{sec:concAndFuture} concludes the paper and proposes avenues for future work.

%% file: background.tex
\section{Background}
\label{sec:background}

This section provides an overview and definition of the concepts used in this work. First, Section~\ref{sec:bg:pose} defines human pose estimation systems, which the proposed framework tests, then Section~\ref{sec:bg:MLTesting} defines challenges specific to testing Machine Learning (ML) systems, and Section~\ref{sec:bg:metamorphic} explains the idea of metamorphic testing, the basis of the proposed framework. 

\subsection{Pose Estimation}
\label{sec:bg:pose}

Pose estimation is the task of estimating the locations of different landmarks on a subject from images or video frames, with the overall aim of providing an overview of their pose. This task is important to a wide variety of fields including human activity recognition~\cite{luvizon20182d}, sign language recognition~\cite{holmes2023scarcity}, and sports analytics~\cite{martin2021automated}. It is typically solved using DL-based regression algorithms which estimate the coordinates of each body part. One of the most common open-source pose estimation frameworks available is MediaPipe's Pose Landmark Detection system \cite{mediapipePose}, which is often incorporated directly into larger ML frameworks for a variety of tasks. This is typically done without fine-tuning, as this model is trained on a large and diverse set of people -- often a far larger number of individuals than would be available for the lower-resource tasks on which this model is typically applied. It is thus important that these pose estimation systems be tested under different conditions when integrated into different systems that target different use cases and thus will provide images of different natures (e.g., dynamic applications will feed the pose estimation system images that are more blurry).

\subsection{Challenges in Testing ML-based Systems}
\label{sec:bg:MLTesting}
Deep Learning (DL) research has gained enormous momentum over the last decade, with a majority of the development in this area being centred around computer vision. However, despite their success, these deep architectures have also demonstrated harmful decision-making~\cite{birhane2024dark, birhane2022auditing,weidinger2021ethical}, bias~\cite{raji2020saving} and a lack of ``understanding" of common-sense concepts such as basic spatial relations~\cite{hoehing2023s, liu2023visual}. These issues are made all the more challenging to identify given that DL methods lack interpretability, with the steps that lead to a particular decision often being unclear. This lack of transparency means that it is crucial that these systems are systematically tested and their results compared to their expected behaviour in order to understand their sensitivities, along with the conditions under which they can be expected to behave correctly. 

Another aspect that complicates the testing of complex systems is that they often present what is known as the \textit{Oracle Problem}~\cite{barr2014oracle}, which describes a scenario where the correct output of the system under test is not known. Sometimes there are no automated ways to compute this oracle, or the act of querying this oracle can be prohibitively expensive. In the case of pose estimation, labelling keypoints in videos is extremely time consuming, making it expensive to collect this form of ground truth data for many applications. This complicates the testing of these systems and requires techniques that can assess the correct functioning of the system without needing an oracle for the correctness of a single execution of the system.

Though testing techniques for computer vision-based classification systems have been extensively explored~\cite{ma2018deepmutation, dwarakanath2018identifying, ma2018deepgauge}, less attention has been given to more complex computer vision tasks such as pose estimation. For example, while MediaPipe does evaluate its pose estimation framework for bias related to the demographics of individuals -- an admirable activity -- this analysis is expensive as it requires ground truth information. Additionally, though their evaluation states that degradation can be expected as video quality and lighting gets worse, a comprehensive testing procedure is not outlined. For many applications, it is crucial that we understand the \textit{specific} parameters within which a system can be expected to operate accurately, especially where mistakes are costly.  There has been some work in this area~\cite{hand-pose}

\subsection{Metamorphic testing}
\label{sec:bg:metamorphic}

\textit{Metamorphic Testing} (\MetTest) has been used to address the oracle problem for both classical programs~\cite{xie2011testing} and ML applications~\cite{dwarakanath2018identifying} and has demonstrated promising results at a relatively low cost. \MetTest relies on \emph{metamorphic rules} -- relations between certain changes to the input of the system (e.g., doubling an integer input) and their effect on its output (e.g., the output should be doubled too) -- to circumvent the oracle problem. If one of these relations is violated for a given input, then the system is not behaving correctly. Note that the violation of the metamorphic rule can be verified without knowing what the correct output for either of the inputs is, i.e., without an oracle for the correctness of each output.

This method is well suited for testing pose estimation systems as it does not require ground truth labels, which come from expensive manual annotation of data. Additionally, since the inputs can be modified in numerous ways, it can help understand the boundaries of the input space under which the system performs as expected.